\documentclass[twocolumn,pra,showpacs,superscriptaddress,amssymb,amsmath,amsmath]{revtex4-1}
\usepackage{graphicx}
\usepackage{epstopdf}
\usepackage{bm}
\usepackage{hyperref}
\usepackage{comment}
\usepackage{color}
\usepackage{physics}
\usepackage{amsmath}

\hypersetup{%
   pdfpagemode=None, 
   pdfstartpage=1,
   pdfmenubar=true,
   pdftoolbar=true,
   colorlinks = true,
   linkcolor=blue,
   citecolor=blue,
   urlcolor=blue,
   bookmarksopen=false
 }

\newcommand{\be}{\begin{equation}}
\newcommand{\ee}{\end{equation}}

\newcommand{\oper}[1]{\mathbf{\hat{#1}}}

\begin{document}

\title{Two interacting ultracold molecules in a one-dimensional harmonic trap}

\author{Anna Dawid}
\affiliation{Faculty of Physics,  University of Warsaw, Pasteura 5, 02-093 Warsaw, Poland}
\author{Maciej Lewenstein}
\affiliation{ICFO-Institut de Ci\`encies Fot\`oniques, The Barcelona Institute of Science and Technology, 08860 Castelldefels, Spain}
\affiliation{ICREA, Pg.~Llu\'is Campanys 23, 08010 Barcelona, Spain}
\author{Micha\l~Tomza}
\email{michal.tomza@fuw.edu.pl}
\affiliation{Faculty of Physics,  University of Warsaw, Pasteura 5, 02-093 Warsaw, Poland}
\date{\today}

\begin{abstract}

We investigate the properties of two interacting ultracold polar molecules described as distinguishable quantum rigid rotors,  trapped in a one-dimensional harmonic potential. The molecules interact via a multichannel two-body contact potential, incorporating the short-range anisotropy of intermolecular interactions including dipole-dipole interaction. The impact of external electric and magnetic fields resulting in Stark and Zeeman shifts of molecular rovibrational states is also investigated. Energy spectra and eigenstates are calculated by means of the exact diagonalization. The importance and interplay of the molecular rotational structure, anisotropic interactions, spin-rotation coupling, electric and magnetic fields, and harmonic trapping potential are examined in detail, and compared to the system of two harmonically trapped distinguishable atoms. The presented model and results may provide microscopic parameters for molecular many-body Hamiltonians, and may be useful for the development of bottom-up molecule-by-molecule assembled molecular quantum simulators.

\end{abstract}


\maketitle

\section{Introduction}

In the last decades an unprecedented control over quantum atomic systems has been achieved at ultralow temperatures. Quantum gases of ultracold atoms in traps have allowed for groundbreaking experiments~\cite{BlochRMP08}. {One-,} two-, and three-dimensional (1D, 2D, and 3D, respectively) systems can now be prepared, manipulated, and measured with great accuracy~\cite{GrossScience17}. Ultracold atoms in traps are especially useful for quantum simulation of various models of many-body physics, and a plethora of quantum phenomena have been investigated and understood~\cite{book:Lewenstein12}. Bose-Einstein condensates, degenerate Fermi gases, quantum phase transitions, and spin models have been realized. In the last years, 1D systems have attracted significant attention~\cite{WeissScience04,ParedesNature04,HallerScience09,PaganoNP14}, due to the important role played by quantum fluctuations~\cite{Giamarchi}.

Recently, the deterministic preparation of tunable few-fermion systems with complete control over the number of particles and their quantum state~\cite{SerwaneScience11} has opened the way towards quantum simulation of strongly correlated few-body systems. The fermionization of two distinguishable fermions~\cite{ZurnPRL12}, formation of a Fermi sea~\cite{Wenz2013}, pairing in few-fermion systems~\cite{ZurnPRL13}, antiferromagnetic Heisenberg spin chains~\cite{MurmannPRL15b}, and two fermions~\cite{MurmannPRL15a} or bosons~\cite{KaufmanScience14,KaufmanNature16} in a double well have been experimentally investigated in one dimension. On the other hand, the atom-by-atom assembling of defect-free 1D and 2D cold atom arrays has also been realized~\cite{EndresScience16,BarredoScience16}. In this way the production of fully controllable synthetic quantum matter can be achieved using both top-down and bottom-up approaches.

Experimental possibilities have motivated intensive theoretical studies of few-body atomic systems. The analytical solution is known  for the general case of two atoms in a harmonic trap interacting via contact~\cite{BuschFP98} or finite-range soft-core~\cite{KoscikSR18} potential. Recently, energy spectra of harmonically trapped two-atom systems with spin-orbit coupling have been investigated in one dimension~\cite{GuanJPB14} and three dimensions~\cite{SchillaciPRA15,YinPRA14}. Low-energy states of two atoms with a dipole moment in a 3D harmonic trap have also been investigated~\cite{OldziejewskiEPL16,Gorecki2016,GoreckiEPL17}. Systems of several fermions~\cite{SowinskiPRA13,Volosniev2014,DeuretzbacherPRA14,GriningPRA15,GriningNJP15,LevinseneSA15,PecakPRA16,PecakNJP16} or several bosons~\cite{DeuretzbacherPRL08,GarciaNJP14,MassignanPRL15,DehkharghaniSR15,YangPRA16} in a 1D harmonic trap have been studied using various analytical and numerical approaches. 

Ultracold molecules have a much richer internal structure as compared to atoms~\cite{CarrNJP09}. This includes rotational and vibrational levels together with possible permanent electric dipole moment. Therefore, one can expect the few-body physics with molecules to be at least as interesting and rich as with atoms. Ultracold high phase-space-density gases of polar molecules in their absolute rovibrational ground state have already been produced~\cite{NiScience08,TakekoshiPRL14,MolonyPRL14,ParkPRL15,GuoPRL16} and allowed for groundbreaking experiments on controlled chemical reactions~\cite{OspelkausScience10,NiNature10,MirandaNatPhys11,McDonaldNature16}. An unprecedented
control over ultracold molecular collisions has been achieved by selecting molecules' internal states and by tuning dipolar collisions with an external electric field in a reduced dimensionality~\cite{QuemenerCR12,LiPRL08,MicheliPRL10,QuemenerPRA11,QuemenerPRA15,TomzaPRL15,IdziaszekNJP15}. 
Ultracold polar molecules have been also produced in an optical lattice~\cite{ChotiaPRL12}, and dipolar spin-exchange interactions between lattice-confined polar molecules have been observed~\cite{BoNature13}, opening the way towards quantum simulations with molecules~\cite{MicheliNatPhys06}. Recently, the first step towards atom-by-atom assembled few-body molecular systems has been taken~\cite{Liu2017} and optical tweezers have been used to assemble and control molecules at the single-particle level~\cite{LiuScience18}.

All the above developments pave the way towards the realization and application of a bottom-up molecule-by-molecule assembled molecular quantum simulator. Here we investigate a fundamental building block of such a simulator, that is, two interacting polar molecules effectively trapped in a one-dimensional harmonic potential. In our model we describe molecules as distinguishable rigid quantum rotors, which interact in a one-dimensional harmonic trap via a multichannel two-body contact potential incorporating short-range isotropic and anisotropic intermolecular interactions. We analyze in detail the properties of such systems including the interplay of the molecular rotational structure, anisotropic interactions, spin-rotation coupling, external electric and magnetic fields, and harmonic trapping potential. Energy spectra and eigenstates are calculated by means of the exact diagonalization. Our calculations may be considered as a microscopic model for the on-site interaction of the molecular multichannel Hubbard Hamiltonian~\cite{DocajPRL16,WallPRA17a,WallPRA17b} and may provide underlying parameters for effective molecular many-body Hamiltonians.

We show that the anisotropic intermolecular interaction brings states with higher total rotational angular momenta to lower energies such that the absolute ground state of the molecular system can have total angular momentum larger than zero and be degenerate.
Such systems may potentially be useful for realizing quantum simulators of exotic spin models. A strong anisotropic intermolecular interaction can induce the emergence of the molecular equivalent of the atomic super-Tonks-Girardeau limit but, at the same time, the importance of the anisotropic intermolecular interaction is reduced in the limit of a very strong isotropic interaction. Magnetic and electric fields induce a high density of states and a large number of avoided crossings, which can be used to control a system's properties. No signatures of quantum chaotic behavior in energy spectra are, however, found.

The plan of this paper is as follows. Section~\ref{sec:theory} describes the used theoretical model and its range of applicability. Section~\ref{sec:results} presents and discusses the numerical results and physical implications of our findings. Section~\ref{sec:summary} summarizes our paper and presents future possible applications and extensions.

\section{Theoretical model}
\label{sec:theory}

\subsection{Hamiltonian}

\begin{figure}[tb!]
\begin{center}
\includegraphics[width=\columnwidth]{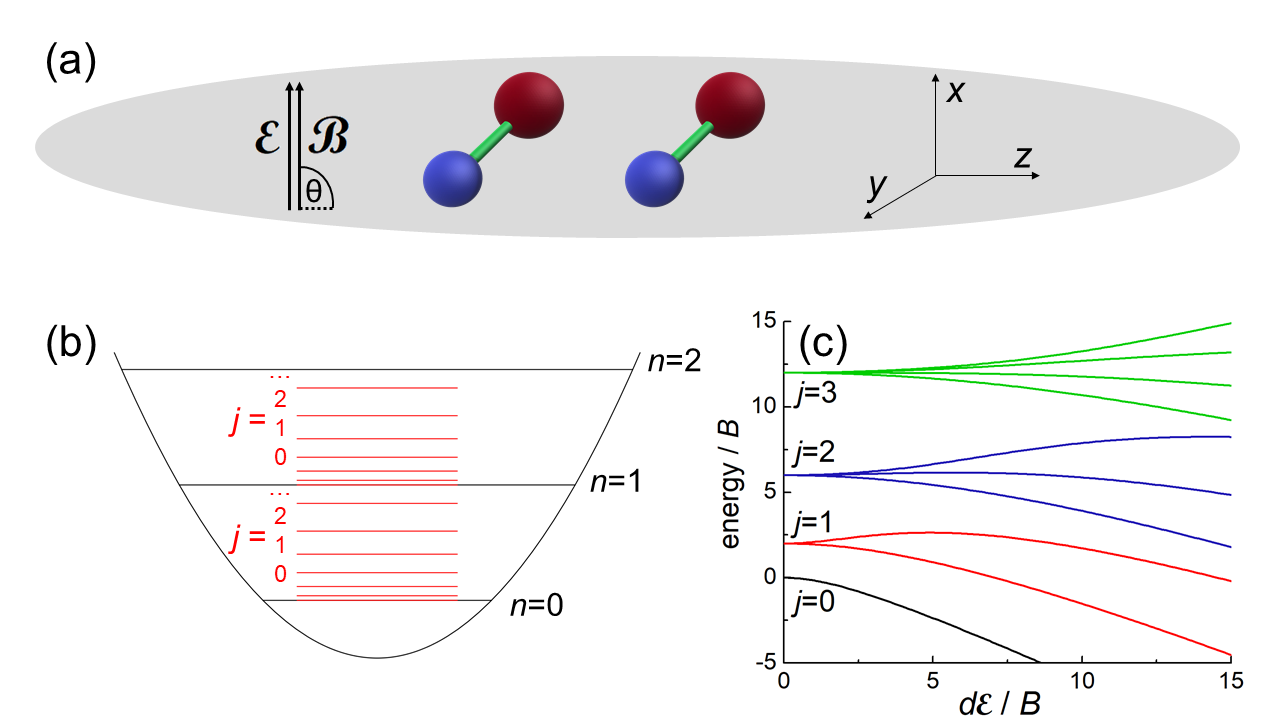}
\end{center}
\caption{Schematic representation of the investigated system and its features:  (a) Illustration of two diatomic polar molecules in a 1D trap in external electric $\bm{\mathcal{E}}$ and magnetic $\bm{\mathcal{B}}$ fields. (b) The energy spectrum of a rotating molecule in a 1D harmonic trap. (c) The energy of a polar molecule in an electric field (Stark's effect).}
\label{fig:model}
\end{figure}

We consider two molecules bound to move along one dimension due to the presence of a strong transverse confinement, as schematically shown in Fig.~\ref{fig:model}(a). Along the axial direction the molecules are further confined by a harmonic potential of frequency $\omega$. The molecules are described within the rigid rotor approximation and they interact by means of a model multichannel two-body contact potential. We assume that the strong transverse confinement does not affect the internal rotation of molecules. This is valid when the size of molecules $R_e$ is much smaller than the transverse harmonic oscillator characteristic length $a_\perp=\sqrt{\hbar/(m\omega_\perp)}$, where $m$ is the mass of molecules and $\omega_\perp$ is the transverse harmonic confinement frequency. At the same time, we assume that the transverse harmonic confinement is much stronger that the axial one ($\omega_\perp\gg\omega$), and no excitation of transverse motion is energetically allowed. We also assume that the confinement does not affect the short-range intermolecular interaction and dynamics. This is valid when the range of chemical intermolecular interactions $R_\mathrm{vdW}$ is much smaller than the harmonic oscillator characteristic lengths $a_\perp$ and $a_{\text{ho}}=\sqrt{\hbar/(m\omega)}$. 
For atomic systems, the effective 1D behavior is observed for elongated harmonic traps with  $\omega_\perp/\omega\geq 10$~\cite{IdziaszekPRA06,SerwaneScience11}.
We neglect possible losses due to inelastic collisions (vibrational relaxation), or chemical reactions, but they can potentially be incorporated in our model within the complex contact interaction potential formalism~\cite{JachymskiPRL13,JankowskaPRA16}.

The generic Hamiltonian describing two interacting polar and paramagnetic molecules in a one-dimensional trap is of the form
\begin{equation}\label{eq:Ham}
\hat{H}=\hat{H}_\mathrm{trap}+\hat{H}_\mathrm{mol}+\hat{H}_\mathrm{field}+\hat{H}_\mathrm{int}\,,
\end{equation}
where $\hat{H}_\mathrm{trap}$ describes the motion of molecules in a trap, $\hat{H}_\mathrm{mol}$ describes the internal (rotational and spin) structure of molecules, $\hat{H}_\mathrm{field}$ describes the interaction of molecules with external fields, and $\hat{H}_\mathrm{int}$ describes the interaction between molecules.

The Hamiltonian describing two structureless particles in a one-dimensional harmonic trap is
\begin{equation}
\hat{H}_\mathrm{trap}=\sum_{i=1}^2 \frac{{\hat{p}}^2_i}{2m}+\sum_{i=1}^2\frac{1}{2}m\omega z_i^2\,,
\end{equation}
where ${\hat{p}}_i$ and $z_i$ are the linear momentum and position of the $i$-th particle, respectively, $m$ is their mass, and $\omega$ is the trapping frequency. We assume that two molecules have the same mass (consisting of the same atoms) and are in the same vibrational state. We also assume that the trapping frequency does not depend on the rotational states of molecules.  

The Hamiltonian describing the internal structure of two molecules (rigid quantum rotors) with spin is 
\begin{equation}
\hat{H}_\mathrm{mol}=\hat{H}_\mathrm{rot}+\hat{H}_\mathrm{spin-rot}\,,
\end{equation}
where
\begin{equation}
\begin{split}\label{eq:Ham_rot}
\hat{H}_\mathrm{rot}=&\sum_{i=1}^2 B\,\mathbf{\hat{j}}_i^2\,,\\
\hat{H}_\mathrm{spin-rot}=&\sum_{i=1}^2  \gamma\,\mathbf{\hat{s}}_i\cdot\mathbf{\hat{j}}_i \,.
\end{split}
\end{equation}
$\hat{H}_\mathrm{rot}$ stands for the rotational structure and $\hat{H}_\mathrm{spin-rot}$ stands for the spin-rotation coupling. $\mathbf{\hat{j}}_i$ is the $i$-th molecule's rotational angular momentum operator, $B$ is the rotational constant, and $\mathbf{\hat{s}}_i$ is the $i$-th molecule's electronic spin angular momentum operator. The molecular spin-rotation interaction with the coupling constant $\gamma$ is responsible for coupling the molecular intrinsic electric and magnetic dipole moments, because the permanent electric dipole moment is associated with the molecular rotation. We assume the same rotational constants and the same electric and magnetic dipole moments for both molecules. We also assume that the rotational and spin-rotation coupling constants do not depend on the rotational states of molecules. We neglect the nuclear spin in our description, although it can be used to control distinguishability of molecules.

The Hamiltonian describing the interaction with external electric and magnetic fields is
\begin{equation}
\hat{H}_\mathrm{field}=\hat{H}_\mathrm{Stark}+\hat{H}_\mathrm{Zeeman}\,,
\end{equation}
where
\begin{equation}
\begin{split}\label{eq:Ham_field}
\hat{H}_\mathrm{Stark}=&-\sum_{i=1}^2 \mathbf{\hat{d}}_i\cdot\bm{\mathcal{E}}\,, \\
\hat{H}_\mathrm{Zeeman}=& 2\mu_B\sum_{i=1}^2\mathbf{\hat{s}}_i\cdot\bm{\mathcal{B}}\,.
\end{split}
\end{equation}
$\mathbf{\hat{d}}_i$ is the $i$-th molecule's electric dipole moment operator and $\bm{\mathcal{E}}$ and $\bm{\mathcal B}$ are the electric and magnetic fields, which couple with the molecules' electric and magnetic dipole moments, respectively. 
$\hat{H}_\mathrm{Stark}$ results in the Stark effect and $\hat{H}_\mathrm{Zeeman}$ results in the Zeeman effect.
We assume that the electric and magnetic fields are parallel to each other and parallel or perpendicular to the motion of molecules in the trap.  

The Hamiltonian describing the interaction between molecules is
\begin{equation}\label{eq:Hamint}
\hat{H}_\mathrm{int}=\hat{H}_\mathrm{iso}+\hat{H}_\mathrm{aniso}\,,
\end{equation}
where we distinguish two parts: the isotropic one $\hat{H}_\mathrm{iso}$ and anisotropic one $\hat{H}_\mathrm{aniso}$. The isotropic part of the interaction potential is of the same nature as the spherically symmetric interaction potential between alkali-metal atoms in the electronic ground state. The anisotropic part of the interaction potential results from the existence of molecular internal structure and orientation dependence of intermolecular interactions~\cite{QuemenerPRA11,QuemenerPRA15}. It is responsible for the transfer of the internal rotational angular momenta between interacting molecules. Neglecting the anisotropic part of the interaction potential restores results known for two interacting atoms in a harmonic trap. 

The intermolecular interaction potential may in general depend on all the internal degrees of freedom and the relative orientation of interacting molecules. This dependence is responsible for effective mixing and exchanging different angular momenta present in interacting molecules and the relative motion during molecular collisions~\cite{HutsonARPC90,QuemenerPRA11,QuemenerPRA15}. Here we propose and employ a model multichannel two-body contact interaction potential to account effectively for the coupling of molecular rotational angular momenta during ultracold molecular collisions. The contact interaction potential is commonly used in ultracold physics as a very successful approximation to describe atom-atom interactions of the short-range van der Waals character~\cite{OlshaniiPRL98}. Similar performance of this approximation can be expected while applied to anisotropic intermolecular interactions of the short-range van der Waals nature. Additionally, in reduced dimensionality, even long-range interactions can be approximated effectively by short-range ones~\cite{HerrickPRA75}.

The isotropic part of the intermolecular interaction potential is 
\begin{equation}\label{eq:Hamiso}
\hat{H}_\mathrm{iso}=\sum_\alpha g_\alpha\delta(z_1-z_2){\hat{P}_{\alpha}}\,,
\end{equation}
where $g_\alpha$ is the strength of the isotropic interaction for channel $\alpha$, $\delta(z)$ is the Dirac delta function imposing assumed contact-type interaction, and $\hat{P}_{\alpha}$ is the projection operator
\begin{equation}\label{eq:Hamaniso}
\hat{P}_{\alpha}= | \alpha \rangle\langle \alpha |\,,
\end{equation}
where $|\alpha\rangle$ is a basis set function (channel) describing all degrees of freedom of the system except the intermolecular distance. 

The anisotropic part of the intermolecular interaction potential is 
\begin{equation}
\hat{H}_\mathrm{aniso}=\sum_{\alpha\neq\alpha'} g_{\alpha\alpha'}\delta(z_1-z_2){\hat{P}_{\alpha\alpha'}}\,,
\end{equation}
where $g_{\alpha\alpha'}$ is the strength of the anisotropic interaction between channels $\alpha$ and $\alpha'$, and $\hat{P}_{\alpha\alpha'}$ is of the form
\begin{equation}
\hat{P}_{\alpha\alpha'}= | \alpha \rangle\langle \alpha' | + | \alpha ' \rangle\langle \alpha |\,.
\end{equation}
Different models of intermolecular interactions can be represented by imposing different forms of $g_{\alpha}$ and $g_{\alpha\alpha'}$.

The isotropic and anisotropic interaction strengths $g_{\alpha}$ and $g_{\alpha\alpha'}$ result from specific values of scattering lengths and chemical short-range intermolecular interaction potentials. In principle, they can be controlled by means of magnetic or optical Feshbach resonances, as well as by changing chemical composition or vibrational states of molecules. 

The general form of the interaction potential of Eq.~\eqref{eq:Hamint} can capture all types of intermolecular interactions. Nevertheless, due to the special importance of the intermolecular dipole-dipole interaction in ultracold physics, we consider this interaction separately. In three dimensions, it is described by the following Hamiltonian
\begin{equation}
\hat{H}_\mathrm{dip}^\text{3D} =\frac{ \oper{d}_1 \cdot \oper{d}_2 - 3 (\oper{d}_1 \cdot \mathbf{e}_r) (\oper{d}_2\cdot \mathbf{e}_r)}{r^3} \,,
\end{equation}
where $r$ and $\mathbf{e}_r$ are the distance and versor connecting two molecules, respectively. By restricting the motion to one dimension, $\mathbf{e}_r = \mathbf{e}_z$, and assuming the contact form of the interaction, $1/z^3\to\delta(z)$, which is the exact result for polarized dipoles in one dimension~\cite{SinhaPRL07,DeuretzbacherPRA10}, we arrive at the effective Hamiltonian, which we use in the present paper
\begin{equation}\label{eq:Hamdip}
\hat{H}_\mathrm{dip} =  - \delta (z_1-z_2) \left(2 \hat{d}_{1,0} \hat{d}_{2,0} + \hat{d}_{1,1} \hat{d}_{2,-1} + \hat{d}_{1,-1} \hat{d}_{2,1} \right)\,,
\end{equation}
where $\hat{d}_{i,q}\equiv \mathbf{e}_q \cdot \oper{d}_i$ are the spherical components of the projection of the dipole operator of the $i$-th molecule onto the space-fixed frame $\mathbf{e}_q$ in spherical coordinates. We approximate the long-range character of the dipole-dipole interaction by the contact potential to simplify our model and analysis. This interaction limited to one dimension does not conserve the total rotational angular momentum.

\subsection{Technicalities of exact diagonalization}

It is possible to separate the center-of-mass and relative motions in the Hamiltonian of Eq.~\eqref{eq:Ham} by introducing new coordinates $Z=\frac{1}{\sqrt{2}}(z_1+z_2)$ and $z=\frac{1}{\sqrt{2}}(z_1-z_2)$, and related momenta $\hat{P}$ and $\hat{p}$. Thanks to unconventional factors of $\sqrt{2}$, the effective masses for both types of motion are the same ($M=\mu=m$). The total wave function in new coordinates $|\Phi(Z,z)\rangle=|\varphi_\text{CM}(Z)\rangle|\Psi_\text{rel}(z)\rangle$ is a product of the wave function for the center-of-mass motion $|\varphi_\text{CM}(Z)\rangle$, which is an eigenstate of the quantum harmonic oscillator Hamiltonian 
\begin{equation}
\hat{H}_\text{CM}=\frac{{\hat{P}}^2}{2m}+\frac{1}{2}m\omega Z^2\,,
\end{equation}
and the wave function for the relative motion $|\Psi_\text{rel}(z)\rangle$, which is a solution of the Schr\"odinger equation with the following Hamiltonian
\begin{equation}\label{eq:relHam}
\hat{H}_\text{rel}=\frac{{\hat{p}}^2}{2m}+\frac{1}{2}m\omega z^2 +\hat{H}_\mathrm{mol}+\hat{H}_\mathrm{field}+\frac{1}{\sqrt{2}}\hat{H}_\mathrm{int}\,.
\end{equation}
In the rest of the paper we focus on finding eigenstates of the above Hamiltonian, therefore whenever we refer to spectra or wave functions of the system we mean properties related to the relative motion. For convenience, we use units of energy and interaction strength that correspond to $\omega=m=\hbar=1$. This amounts to measuring energies $E$ in units of $\hbar\omega$, lengths in units of the harmonic oscillator characteristic length $a_\text{ho}=\sqrt{\hbar/(m\omega)}$, and the interaction strengths $g_\alpha$ and $g_{\alpha\alpha'}$ in units of $\hbar\omega a_\text{ho}$.

For two polar molecules without spin, we represent the relative motion part of the total wave function in the basis of eigenstates of the one-dimensional harmonic oscillator $|n\rangle$ [$(\frac{{\hat{p}}^2}{2m}+\frac{1}{2}m\omega z^2)|n\rangle=(n+\frac{1}{2})\hbar\omega|n\rangle$] and eigenstates of the total rotational angular momentum operator $|J,M,j_1,j_2\rangle$ [$\hat{\mathbf{J}}^2|J,M,j_1,j_2\rangle=J(J+1)|J,M,j_1,j_2\rangle$ and $\hat{\mathbf{J}}_z|J,M,j_1,j_2\rangle=M|J,M,j_1,j_2\rangle$ with $\hat{\mathbf{J}}=\hat{\mathbf{j}}_1+\hat{\mathbf{j}}_2$]
\begin{equation}\label{eq:psi}
|\Psi_k\rangle=\sum_{\substack{n,J,M,j_1,j_2}}C^k_{\substack{n,J,M,j_1,j_2}}|n\rangle|J,M,j_1,j_2\rangle\,,
\end{equation}
where
\begin{equation}\label{eq:basis}
|J,M,j_1,j_2\rangle=\sum_{m_1,m_2}\langle j_1,m_1,j_2,m_2 |J,M\rangle |j_1,m_1\rangle|j_2,m_2\rangle\,,
\end{equation}
where $\langle j_1,m_1,j_2,m_2 |J,M\rangle$ are Clebsch-Gordan coefficients and $|j_i,m_i\rangle$ are eigenstates of $\hat{\mathbf{j}}_i$.
The symmetries of the Hamiltonian of Eq.~\eqref{eq:relHam} resulting in the conservation of $J$ or $M$ quantum numbers are used to restrict properly the size of the employed basis set.
All possible combinations of basis functions with $n\leq n_\text{max}$, $j_1\leq j_\text{max}$, and $j_2\leq j_\text{max}$ are employed in calculations.
Numerical coefficients $C^k_{\substack{n,J,M,j_1,j_2}}$ for the $k$-th state are calculated by means of the exact diagonalization method.

In the computational basis set (channels) introduced in Eqs.~\eqref{eq:psi} and~\eqref{eq:basis}, the projection operator $\hat{P}_\alpha$ in the isotropic part of the intermolecular interaction potential of Eq.~\eqref{eq:Hamiso} takes the form
\begin{equation}
\hat{P}_\alpha=|J,M,j_1,j_2\rangle\langle J,M,j_1,j_2|\,.
\end{equation}
We assume that the corresponding strengths of the isotropic interaction are independent of molecular internal states and the same for all channels
\begin{equation} 
g_\alpha = g_0\,.
\end{equation}

The coupling operator $\hat{P}_{\alpha\alpha'}$ in the anisotropic part of the intermolecular interaction potential of Eq.~\eqref{eq:Hamaniso} is of the form
\begin{equation}
\hat{P}_{\alpha\alpha'}=|J,M,j_1,j_2\rangle\langle J,M,j'_1,j'_2|+ \text{H.c.}\,,
\end{equation}
where we assume that the anisotropic part of the intermolecular interaction potential does conserve the total rotational angular momentum of two molecules $J,M$. 
The corresponding strength of the anisotropic interaction can be written as 
\begin{equation}
g_{\alpha\alpha'}\equiv g_{JMj_1j_2,JMj'_1j'_2}\,.
\end{equation}
We assume that the anisotropic interaction strengths do not depend on the total angular momentum $J,M$
\begin{equation}
g_{JMj_1j_2,JMj'_1j'_2} = g_{j_1j_2,j'_1j'_2}\,,
\end{equation}
and consider two types of the anisotropic interaction. The first one couples molecular states which differ by $k$ quanta of molecular rotational angular momentum
\begin{equation}
g_{j_1j_2,j_1'j_2'}^{\pm k}= \delta_{j_1,j_1'\pm k}\delta_{j_2,j_2'\mp k} \,g_{\pm k}\,,
\end{equation}
In this notation, the dipole-dipole interaction has non-zero terms only related to $g_0$ and $g_{\pm 2}$.
The second, simplified type allows us only to exchange the rotational angular momenta between two interacting molecules if they differ by $k=|j_1-j_2|$
\begin{equation}
g_{j_1j_2,j_1'j_2'}^{\pm k,\text{ex}} = \delta_{j_1,j_2'}\delta_{j_2,j_1'} \delta_{j_1\pm k,j_2}\,g_{\pm k}^\text{ex}\,.
\end{equation}
The coefficients $g_{\pm k}$ and $g_{\pm k}^\text{ex}$ can depend on $k$. In our model calculations we assume that these coefficients are the largest for $k=1$ and we neglect them for $k>1$ or assume geometric decay with $k$, $g_{\pm k} = {g_{\pm 1}}/{A^{k-1}}$.

When we consider molecules possessing spin, the basis set used in Eq.~\eqref{eq:psi} is augmented by the product of eigenstates $|s_i,m_{s_i}\rangle$ of molecular electronic spin operators $\hat{\mathbf{s}}_i$ resulting in the computational basis of the form
\begin{equation}
|n\rangle|J,M_J,j_1,j_2\rangle |s_1,m_{s_1}\rangle|s_2,m_{s_2}\rangle\,.
\end{equation}
All spin configurations allowed by the symmetry are included in the basis set. The total angular momentum of the system $J_\text{tot},M_\text{tot}$ is then the sum of the total rotational and spin angular momenta. We assume that the intermolecular interaction potential of Eq.~\eqref{eq:Hamint} does not depend on the electronic spin. Matrix elements of the system's Hamiltonian in the employed basis set are provided in Appendix~\ref{sec:app}. 

In all calculations we use basis sets with $n_\text{max}=30$ and $j_\text{max}=8$, unless it is stated otherwise. Without using symmetries, the size of the Hamiltonian matrices would be $10^5-10^6$. If symmetries are employed, the size of the Hamiltonian matrices to be diagonalized is between around $10^3$ and $2\cdot 10^4$, depending on the angular momentum and the presence of external fields and spin structure. The relatively slow convergence of correlated energy calculations for interacting particles in one dimension with the number of used single-particle harmonic oscillator functions has been recently shown~\cite{GriningNJP15,GriningPRA15}. Nevertheless, for the range of weak and intermediate interaction strengths, physically meaningful results can be obtained with used $n_\text{max}$. Additionally, in the present case, we have checked that the convergence of energy calculations with the size of the rotational basis set is much faster and obtained results are close to converged with respect to $j_\text{max}$~\cite{DawidMSc17}.

In the present paper we assume that the rotational constant of molecules is smaller than or equal to the frequency of the trap, $B\leq\omega$. If the rotational constant is much larger than the frequency of the trap, $B\gg\omega$, then the results approach atomic solution and molecular features are less important. To meet the condition $B\leq\omega$, molecules in a weakly bound state (e.g., Feshbach molecules~\cite{JulienneRMP06b}) or the very tight harmonic trap (e.g., a nanoplasmonic one~\cite{ThompsonScience13}) can be used. 

\section{Results}
\label{sec:results}

\subsection{Short-range anisotropy of intermolecular interaction}

\begin{figure}[tb!]
\begin{center}
\includegraphics[width=0.95\columnwidth]{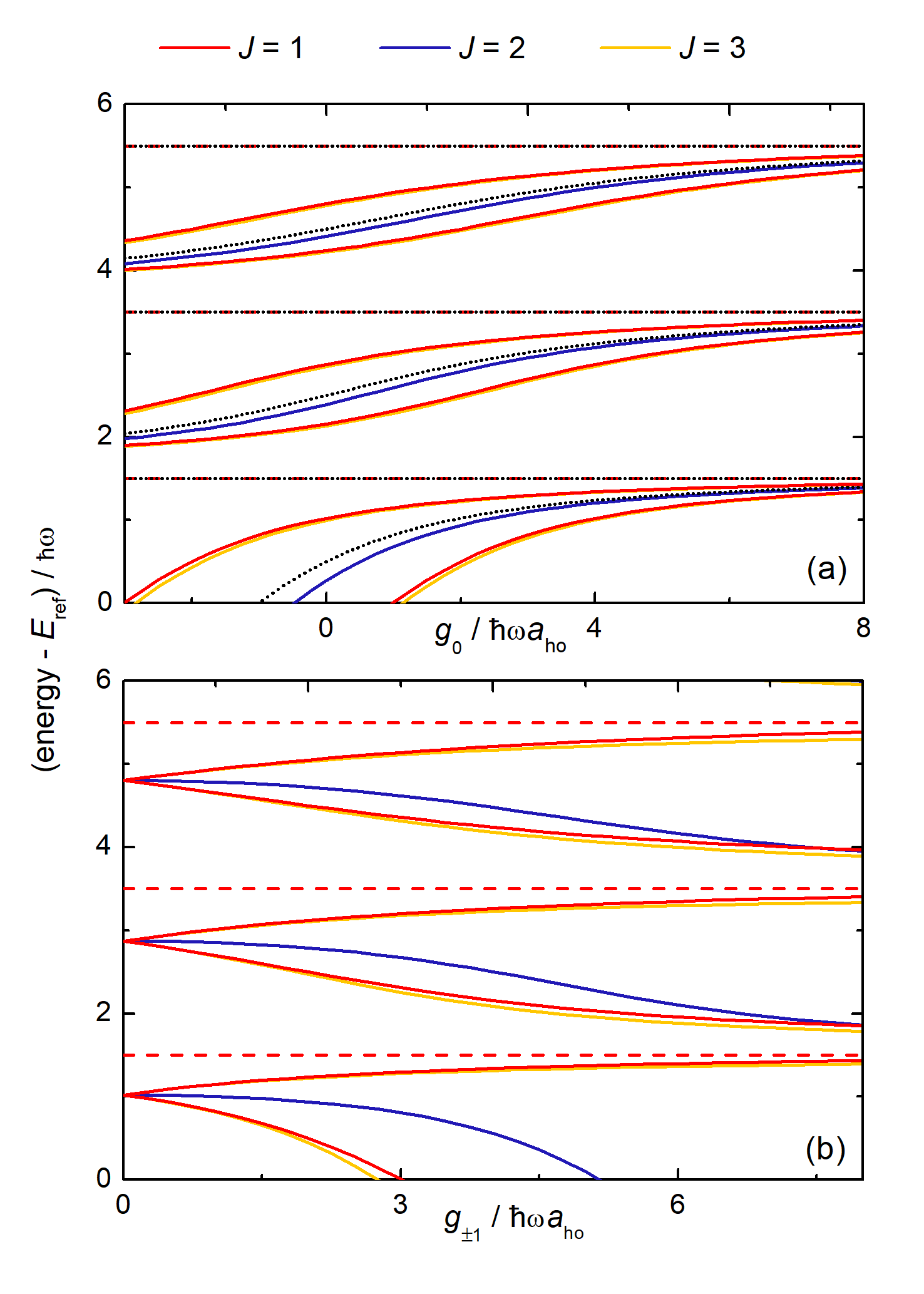}
\end{center}
\caption{Energy spectra of the relative motion for two interacting molecules with the rotational constant $B=10\,\hbar\omega$ in a one-dimensional harmonic trap: (a) as a function of the isotropic interaction strength $g_0$ with the anisotropic interaction strength $g_{\pm 1} = 2$ and (b) as a function of the anisotropic interaction strength $g_{\pm 1}$ with the isotropic interaction strength $g_0 = 2$. The spectra for different total angular momentum $J$ are shifted by the energy of non-interacting systems with this total angular momentum. Solid and dashed lines are for states of even and odd spatial symmetries, respectively. Dotted lines in panel (a) are the result for two interacting atoms.}
\label{fig:largeB}
\end{figure}

\begin{figure}[tb!]
\begin{center}
\includegraphics[width=0.95\columnwidth]{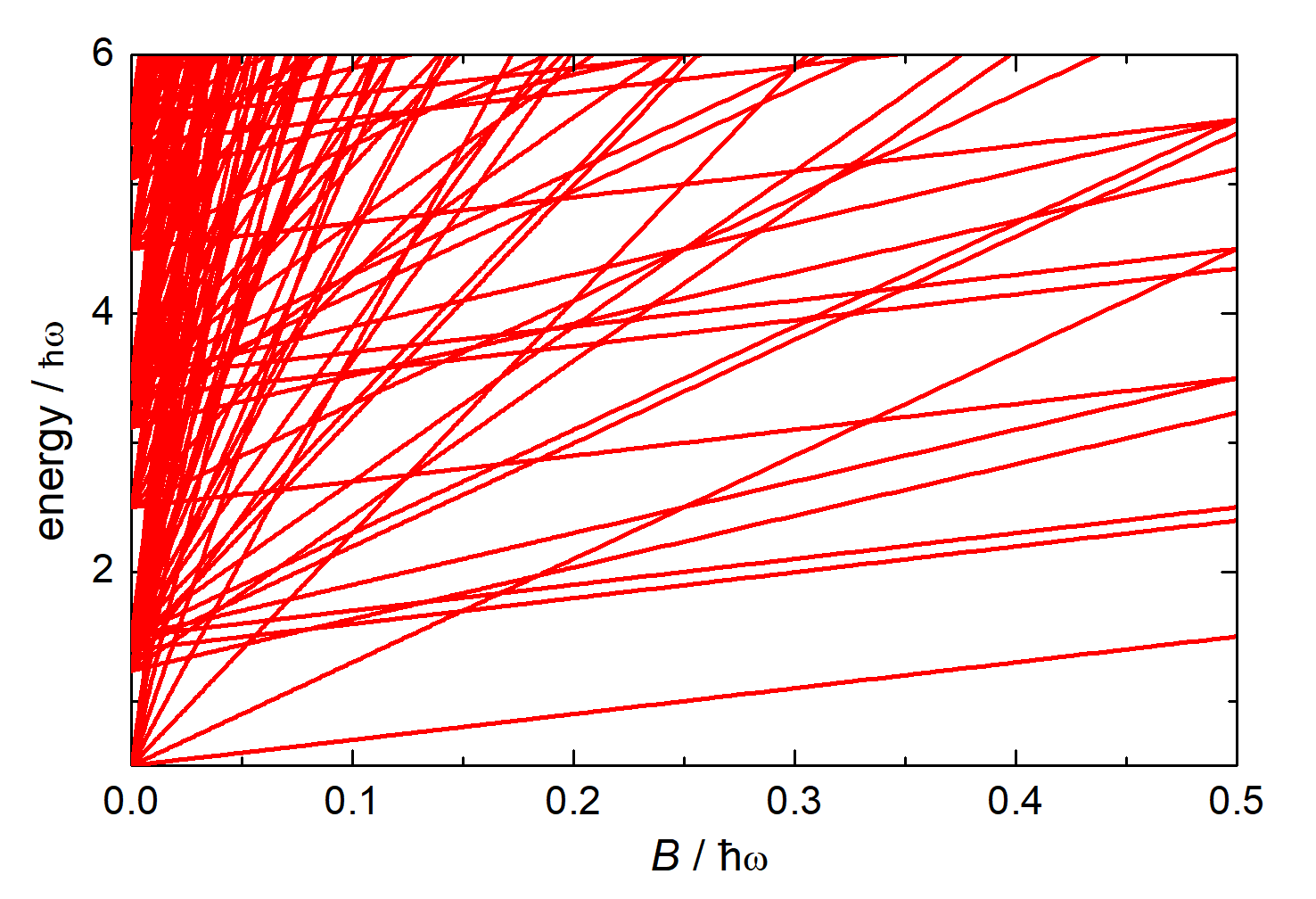}
\end{center}
\caption{Energy spectrum of the relative motion for two interacting molecules in a one-dimensional harmonic trap as a function of their  rotational constant $B$. The isotropic and anisotropic interaction strengths are set at $g_0=g_{\pm 1}=4$.}
\label{fig:EvsrotB}
\end{figure}

Before we focus on systems with small rotational constants $B\leq\omega$, which are the main subject of this paper, we will analyze the impact of the anisotropic interaction on systems with $B \gg \omega$, where the effect of the anisotropic interaction is relatively smaller, but easier to interpret. Fig.~\ref{fig:largeB} presents energy spectra of the relative motion for two interacting molecules with the rotational constant $B=10\,\hbar\omega$ in a one-dimensional harmonic trap. Results for three total angular momenta $J=1, 2$, and $3$ are presented as a function of the isotropic interaction strength $g_0$ with the anisotropic interaction strength $g_{\pm 1}=2$ in panel (a) and as a function of the anisotropic interaction strength $g_{\pm 1}$ with the isotropic interaction strength $g_0=2$ in panel (b). Energy spectra are compared with the known result for two interacting atoms~\cite{BuschFP98}, that is equivalent to the energy spectrum for interacting molecules with $g_{\pm 1}=0$ or with $J=0$. In panel (a), energies of states with $J=2$ are only slightly shifted as compared with the atomic case due to the coupling by the anisotropic interaction with higher-energy states. Instead, there are two energy states for each branch for $J=1$ and $3$. They originate from the fact that, in the presented energy range, those total angular momenta can be constructed from two rotational configurations with $j_1=1,j_2=0$ and $j_1=0,j_2=1$ for $J=1$, and $j_1=2,j_2=1$ and $j_1=1,j_2=2$ for $J=3$, which are coupled by the anisotropic interaction, whereas the lowest state with $J=2$ originates from a single rotational configuration with $j_1=1,j_2=1$.
The emergence of the splitting between two states for $J=1$ and $3$, and the shift for $J=2$ as a function of the anisotropic interaction strength, are presented in panel (b).   
The energy spectra for $J=1$ and $3$ are very similar to each other, because in our model the anisotropy of the intermolecular interaction is assumed to be independent of the total rotational angular momentum.

To choose a rotational constant for further investigations, Fig.~\ref{fig:EvsrotB} presents the energy spectrum of the relative motion for two interacting molecules in a one-dimensional harmonic trap as a function of the molecules' rotational constant $B$ with the isotropic and anisotropic interaction strengths set at $g_0=g_{\pm 1}=4$. For the unphysical regime of $B\ll \omega$, the energy spectrum becomes very dense. Detailed results for $B\gg \omega$ will be presented elsewhere. Therefore, in the rest of the present paper we assume an intermediate value of the rotational constant $B=0.3\,\hbar\omega$. 

\begin{figure*}[tb!]
\begin{center}
\includegraphics[width=0.95\textwidth]{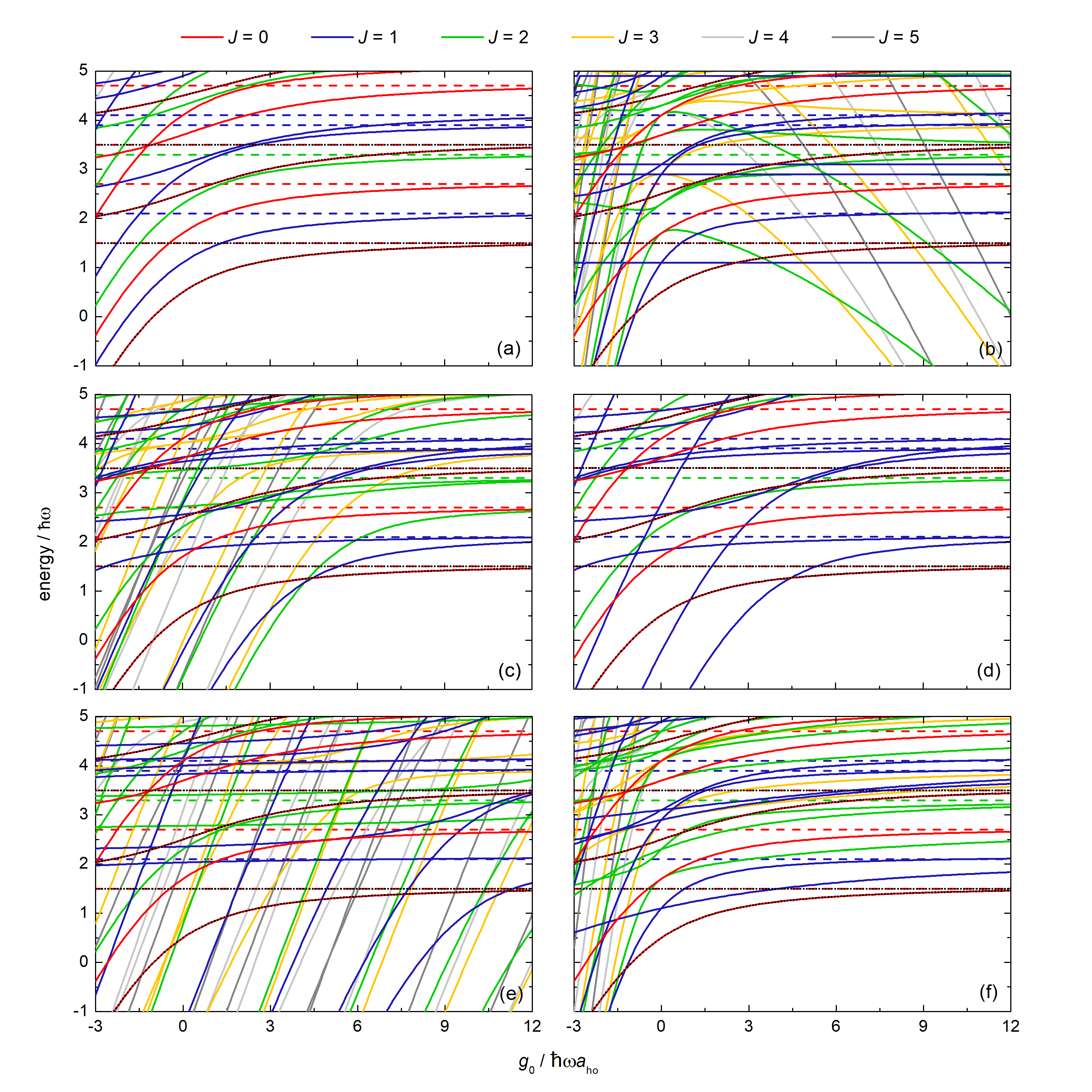}
\end{center}
\caption{Energy spectra of the relative motion for two interacting molecules with the rotational constant $B=0.3\,\hbar\omega$ in a one-dimensional harmonic trap as a function of the isotropic interaction strength $g_0$ for different models of the short-range anisotropy of intermolecular interaction: (a)~$g_{\pm k}=0$, (b)~$g_{\pm 1}=g_0$, (c)~$g_{\pm 1}=4$, (d)~$g_{\pm 1}^\text{ex}=4$, (e)~$g_{\pm 1}=10$, and (f)~$g_{\pm k}={g_0}/{1.5^k}$. Solid and dashed lines are for states of even and odd spatial symmetries. Different colors represent states with different total rotational angular momenta. Dotted lines are for the result for two interacting atoms.}
\label{fig:Evsgjj}
\end{figure*}

\begin{figure}[tb!]
\begin{center}
\includegraphics[width=0.98\columnwidth]{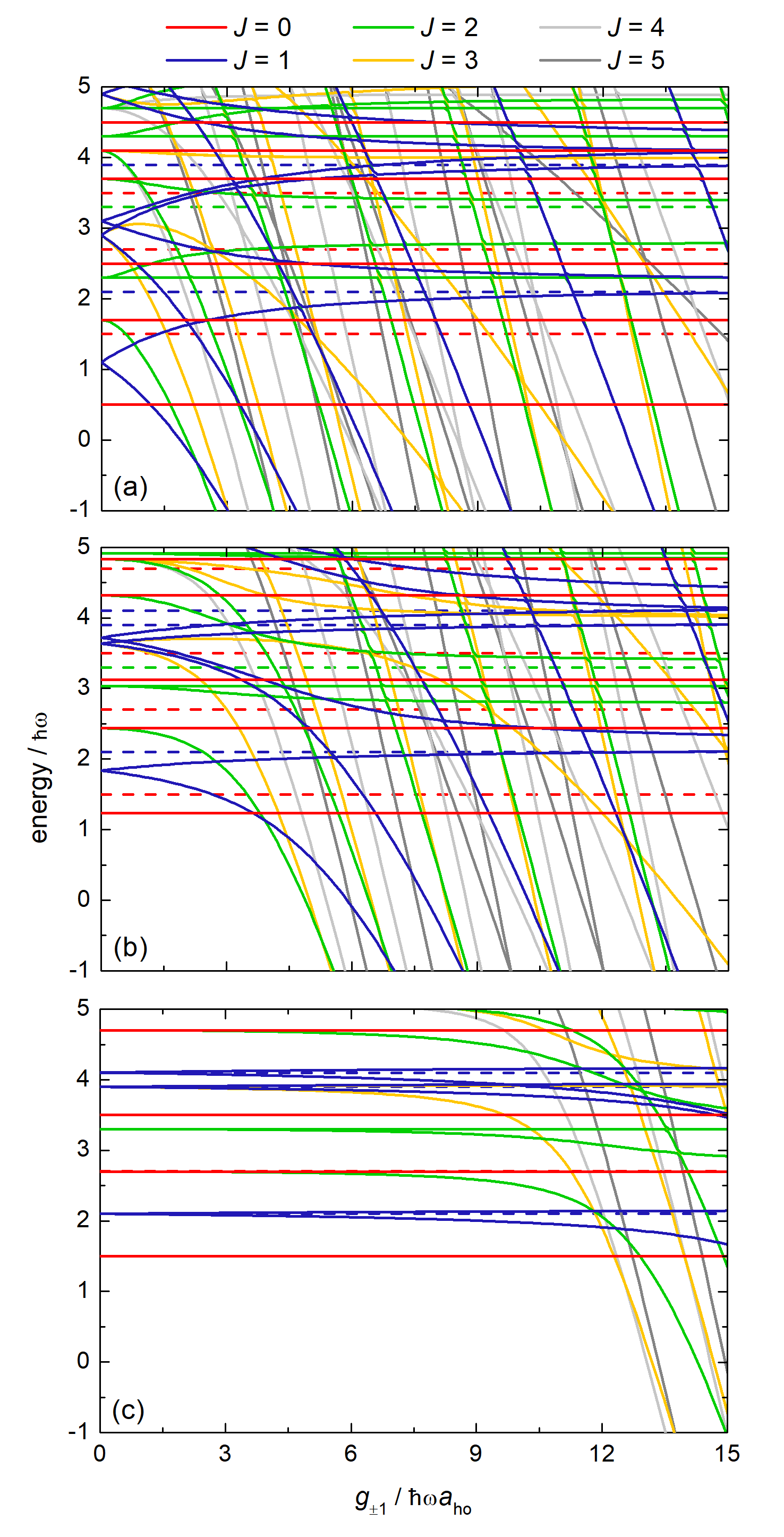}
\end{center}
\caption{Energy spectra of the relative motion for two interacting molecules with the rotational constant $B=0.3\,\hbar\omega$ in a one-dimensional harmonic trap as a function of the anisotropic interaction strength $g_{\pm 1}$ for different isotropic interaction strengths: (a)~$g_0=0$, (b)~$g_0=4$, and (c)~$g_0=17.3$ ($g_0^\text{eff}=\infty$ for a calculation in a finite basis set). Solid and dashed lines are for states of even and odd spatial symmetries. Different colors represent states with different total rotational angular momenta.}
\label{fig:Evsgjk}
\end{figure}

Figure~\ref{fig:Evsgjj} shows the dependence of the energy spectra of the relative motion for two interacting molecules with the rotational constant $B=0.3\,\hbar\omega$ in a one-dimensional harmonic trap on the isotropic interaction strength $g_{0}$ for different models of the anisotropic interaction. The states with the even and odd spatial symmetries are denoted by solid and dashed lines, respectively, and they are compared with the known result for two atoms denoted by the dotted lines. Because of the assumed contact-type interaction potential, the energy of states with odd symmetry does not depend on the intermolecular interactions, similarly as in the atomic case~\cite{BuschFP98}. Additionally, for systems without any rotational angular momentum, $J=0$, $j_1=0$, and $j_2=0$, the assumed form of the anisotropic intermolecular interaction does not affect the system, and energies reduce to the atomic spectrum, in agreement with collisional results obtained with a complete description of intermolecular interactions~\cite{HutsonARPC90}. As a reference, Fig.~\ref{fig:Evsgjj}(a) presents the energy spectrum for the molecular system without any anisotropic interaction, thus it corresponds to the atomic result multiplied and shifted by rotational energies only and it reveals a complex nature of investigated systems resulting from the richer internal structure of molecules as compared to atoms.   

Figures~\ref{fig:Evsgjj}(b)-(f) present the energy spectra with non-zero anisotropic intermolecular interactions in different scenarios. In Fig.~\ref{fig:Evsgjj}(b) we assume that the anisotropic interaction strength is the same as the isotropic one, $g_{\pm 1}=g_0$. Interestingly, in such a case, some energy levels for higher total rotational angular momenta diverge to minus infinity with increasing $g_0$. 
This indicates that molecules form clusters deeply bound by a strong anisotropic interaction, while other levels, which converge to constant energies, can be interpreted as metastable gas-like super-Tonks states~\cite{AstrakharchikPRL05,TempfliNJP08,HallerScience09}. Thus, a strong anisotropic intermolecular interaction can induce the existence of the molecular equivalent of the atomic super-Tonks-Girardeau limit in investigated systems. Specifically, in our numerical tests, we have observed such a behavior for interaction models with $g_{\pm 1}/g_0\geq 1$.


Figures~\ref{fig:Evsgjj}(c) and~\ref{fig:Evsgjj}(e) present the energy spectra for the system with the anisotropic interaction strength set at $g_{\pm 1}=4$ and $10$, respectively. A larger anisotropy leads to a larger distortion of the spectrum as compared to the atomic case and brings states with higher total rotational angular momentum to lower energies. Figure~\ref{fig:Evsgjj}(d) presents the energy spectrum for the simplified version of the anisotropic interaction $g_{\pm 1}^\text{ex}=4$ allowing for exchanging angular momentum only. As expected, the spectrum in this case is distorted only for total angular momenta for which the assumed form of the intermolecular anisotropic interaction affects the system. Finally, Fig.~\ref{fig:Evsgjj}(f) presents the energy spectrum for the anisotropic interaction, which is proportional to the isotropic interaction strength, is non-zero for higher $k$, but decays geometrically with $k$, $g_{\pm k}={g_0}/{1.5^k}$. Since the anisotropic interaction strength in this model is always smaller than the isotropic one, the molecular features in the spectrum are less pronounced, despite the presence of couplings for higher $k$. In general, the anisotropic interactions related to larger $k$ in our model are less important than the leading coupling term $g_{\pm 1}$, because they couple states with increasing energy differences. 

Figure~\ref{fig:Evsgjk} shows the dependence of the energy spectra of the relative motion for two interacting molecules with the rotational constant $B=0.3\,\hbar\omega$ in a one-dimensional harmonic trap on the anisotropic interaction strength $g_{\pm k}$ for different values of the isotropic interaction strength $g_0$. Panels (a), (b), and (c) present results for very small ($g_0=0$), intermediate ($g_0=4$), and very large ($g_0\to\infty$) isotropic interaction strengths, respectively. The impact of the anisotropic intermolecular interaction decreases with increasing the isotropic interaction strength. Especially, in the limit of a very large isotropic interaction strength ($g_0\to\infty$), corresponding to the Tonks-Girardeau limit in atomic systems, the strength of the anisotropic intermolecular interaction has to be tuned to very large values to induce observable effects. It is not surprising, since for large positive (repulsive) values of the isotropic interaction, that is, in the Tonks-Girardeau limit, the interacting particles avoid each other, decreasing their wave functions' overlap and thus decreasing the effect of the short-range anisotropic interaction. For a negative strength of the isotropic interaction interaction, $g_0<0$, the anisotropic interaction affects systems more easily because molecules are attracted to each other. The energy spectra for simplified versions of the anisotropic interaction allowing for exchanging angular momentum only are distorted only for total angular momenta for which the assumed form of the intermolecular anisotropic interaction affects the system.

A common and interesting feature for all investigated models of intermolecular interactions analyzed in Figs.~\ref{fig:Evsgjj} and~\ref{fig:Evsgjk} is that, in the presence of the anisotropic interaction, the absolute ground state of the system can have total angular momentum larger than zero, $J>0$. Such a ground state has a $(2J+1)$ degeneracy that can allow for the realization of interesting many-body Hamiltonians in the limit of many optical lattice sites each occupied by two molecules. The ground state and its degeneracy in such a scenario can be controlled by tuning the anisotropy of the intermolecular interaction.

\begin{figure}[tb!]
\begin{center}
\includegraphics[width=0.98\columnwidth]{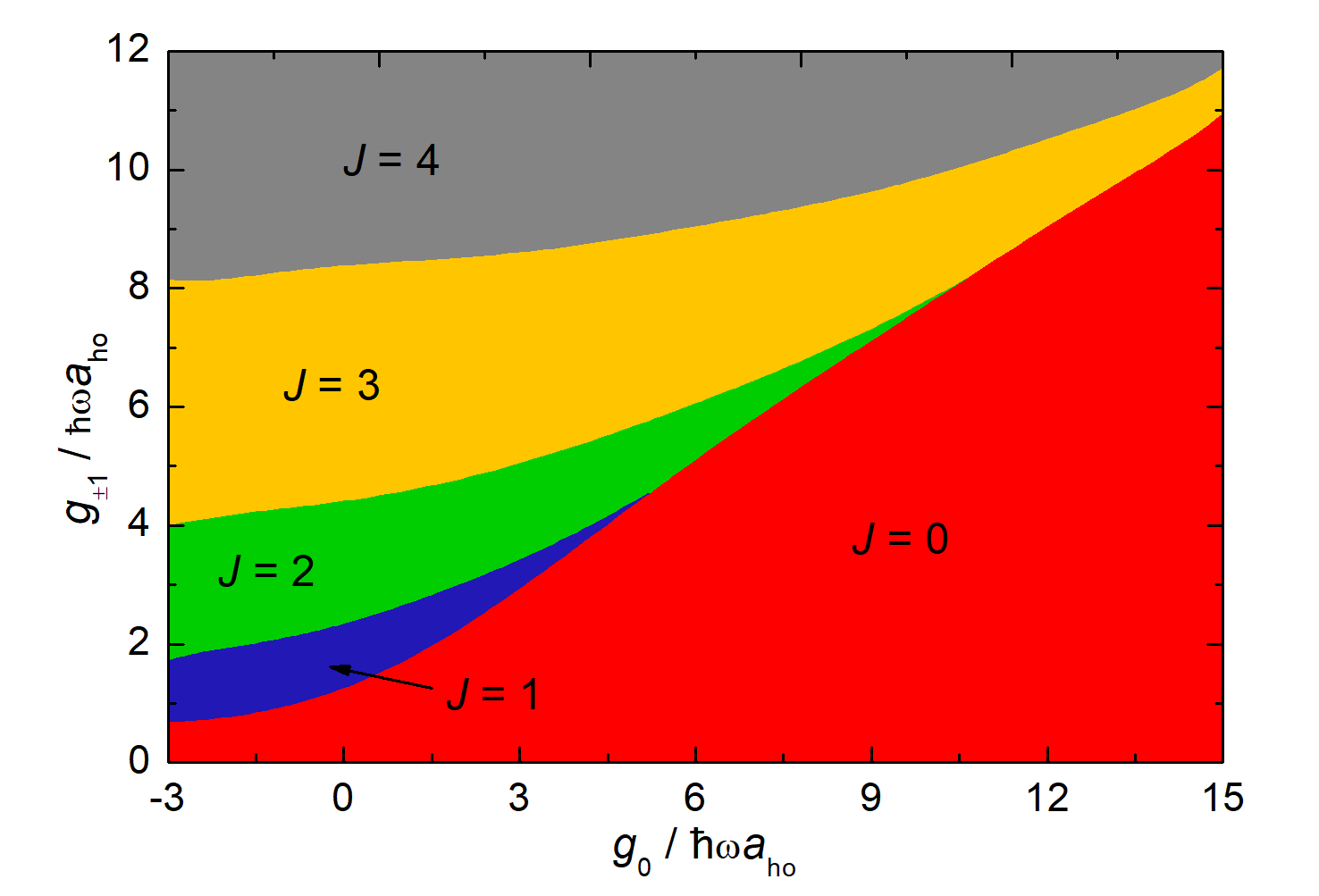}
\end{center}
\caption{Total rotational angular momentum $J$ of the ground state for two interacting molecules with the rotational constant $B=0.3\,\hbar\omega$ in a one-dimensional harmonic trap as a function of the isotropic $g_0$ and anisotropic $g_{\pm 1}$ interaction strengths.}
\label{fig:phase_diagram}
\end{figure}

Figure~\ref{fig:phase_diagram} presents the dependence of the total rotational angular momentum $J$ of the ground state for two interacting molecules with the rotational constant $B=0.3\,\hbar\omega$ in a one-dimensional harmonic trap on the isotropic $g_0$ and anisotropic $g_{\pm 1}$ interaction strengths. This plot clearly shows the interplay of the isotropic and anisotropic intermolecular interactions observed already in Figs.~\ref{fig:Evsgjj} and~\ref{fig:Evsgjk}. In the absence of or for weak anisotropic interactions, the ground state has $J=0$. With increasing strength of the anisotropic interaction, the ground state has increasingly higher total rotational angular momentum. For small isotropic interaction strengths it is easier to induce incrementally higher total rotational angular momentum in the ground state. For large isotropic interaction strengths, much larger anisotropic interaction strengths are needed to induce the ground state with higher $J$, and ground states with higher $J$ neighbor one with $J=0$. 

\begin{figure}[tb!]
\begin{center}
\includegraphics[width=0.98\columnwidth]{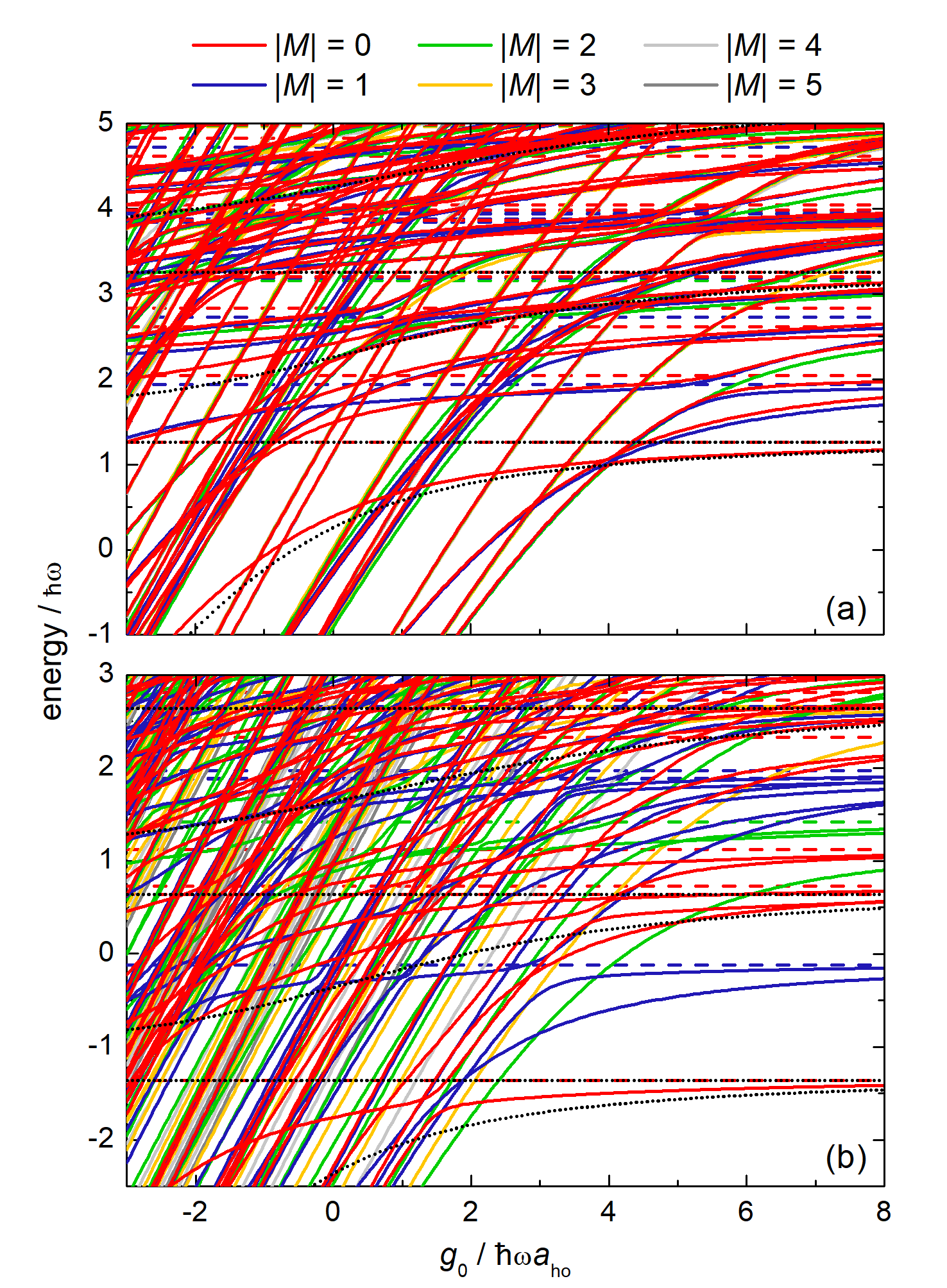}
\end{center}
\caption{Energy spectra of the relative motion for two interacting molecules with the rotational constant $B=0.3\,\hbar\omega$ in a one-dimensional harmonic trap in an external static electric field as a function of the isotropic interaction strength $g_0$ with the anisotropic interaction strength $g_{\pm 1}=4$ and the electric field strengths. (a) $d \mathcal{E}=0.5\,\hbar\omega$ and (b) $d\mathcal{E}=2.5\,\hbar\omega$. Solid and dashed lines are for states of even and odd spatial symmetries. Different colors represent states with different projections $|M|$ of the total rotational angular momentum along the field. Dotted lines are for the result for two interacting atoms shifted by the energy of two non-interacting polar molecules.}
\label{fig:Evsgjj+elf}
\end{figure}
\begin{figure}[tb!]
\begin{center}
\includegraphics[width=0.98\columnwidth]{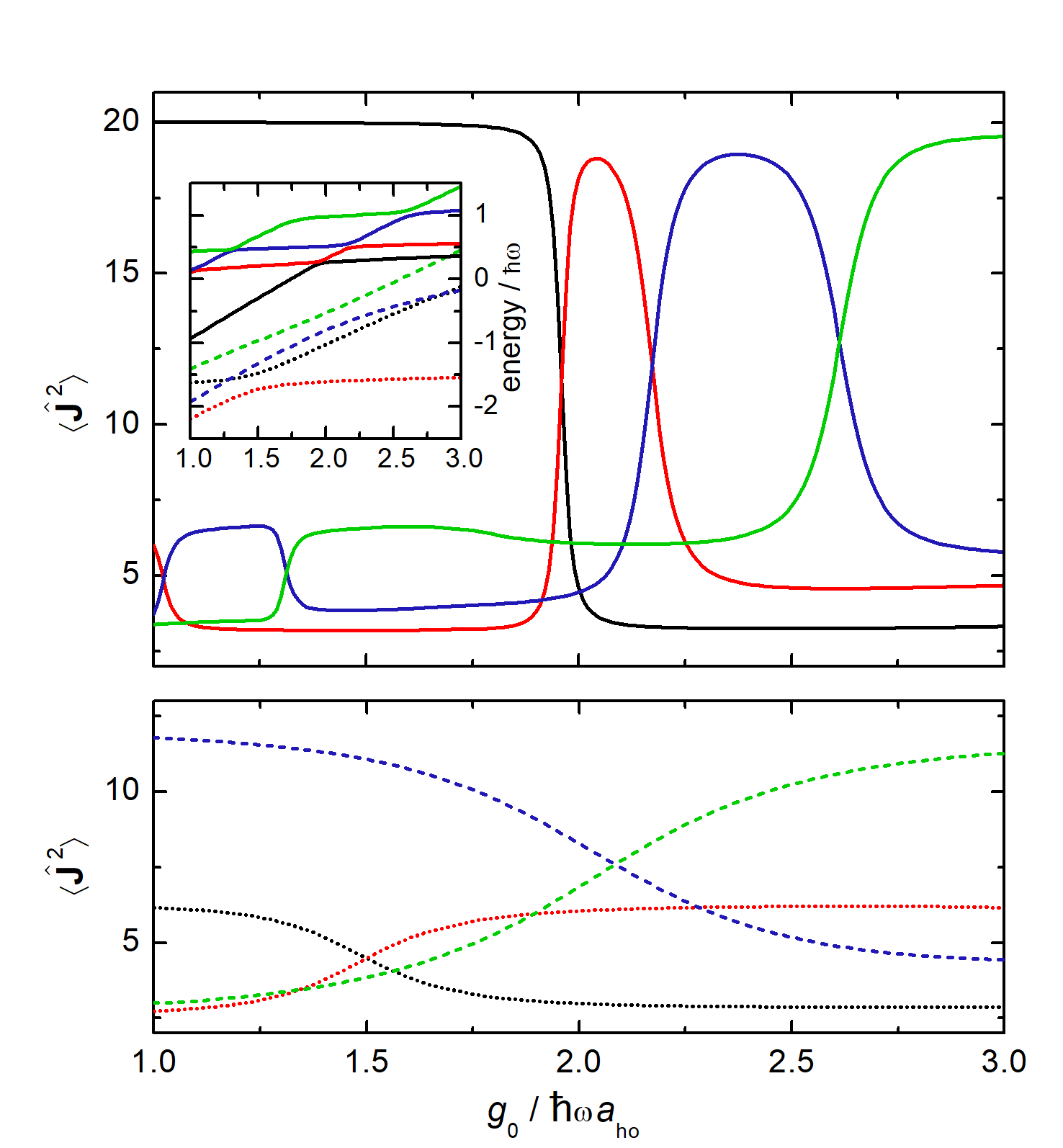}
\end{center}
\caption{Mean values of the square of the total rotational angular momentum operator $\hat{\mathbf{J}}^2$ for selected eigenstates of two interacting molecules with the rotational constant $B=0.3\,\hbar\omega$ in a one-dimensional harmonic trap in an external static electric field as a function of the isotropic interaction strength $g_0$ with the anisotropic interaction strength $g_{\pm 1}=4$ and electric field strength $d\mathcal{E}=2.5\,\hbar\omega$. Solid, dashed, and dotted lines represent states with different
projections $|M|$ of the total rotational angular momentum along the field. The inset shows the energy spectrum of analyzed eigenstates using the same color code.}
\label{fig:J2vsg}
\end{figure}

In the absence of external electric or magnetic fields, the total rotational angular momentum is a conserved quantity ($J$ is a good quantum number). Therefore, it is not possible to drive the system between ground states with different total rotational angular momenta by simply tuning systems' parameters. However, transitions involving photon absorption or emission can potentially be used to reach the ground state with higher total rotational angular momentum after changing the system's parameters. Adiabatic evolution between different ground states can, however, be possible if an external electric field is applied to couple states with different $J$.

\subsection{Impact of external electric field}
\label{sec:el_field}

\begin{figure*}[tb!]
\begin{center}
\includegraphics[width=0.98\textwidth]{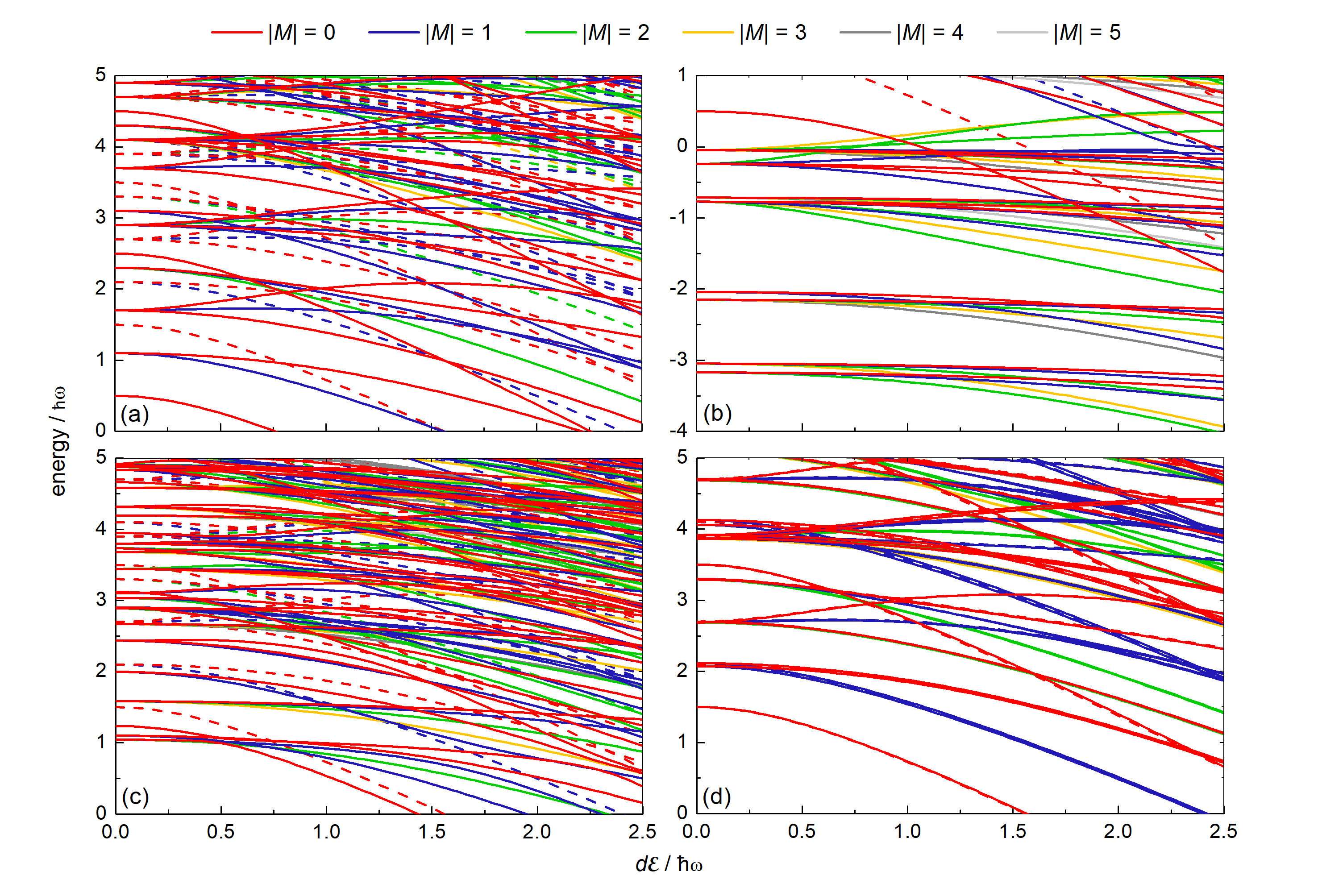}
\end{center}
\caption{Energy spectra of the relative motion for two interacting molecules with the rotational constant $B=0.3\,\hbar\omega$ in a one-dimensional harmonic trap in an external static electric field as a function of the electric field strength $d \mathcal{E}$ for different models of the isotropic and anisotropic short-range interactions: (a)~$g_0=0$, $g_{\pm 1}=0$; (b)~$g_0=0$, $g_{\pm 1}=4$; (c)~$g_0=4$, $g_{\pm 1}=4$; and (d)~$g_0=17.3$ ($g_0^\text{eff}=\infty$ for a calculation in a finite basis set), $g_{\pm 1}=4$. 
Solid and dashed lines are for states of even and odd spatial symmetries. Different colors represent states with different projections $|M|$ of the total rotational angular momentum along the field.}
\label{fig:Evself}
\end{figure*}

If considered molecules are heteronuclear and posses a permanent electric dipole moment, then a static electric field can be used as a knob to control their interactions and dynamics in a trap via Stark effect~\cite{LemeshkoMP13}. An electric field couples and mixes states with different total rotational angular momenta $J$ and removes the degeneracy of states with different $|M|$. The energy spectrum of a single polar molecule in an electric field is shown in Fig.~\ref{fig:model}(c) as a reference.

Figure~\ref{fig:Evsgjj+elf} presents energy spectra of the relative motion for two interacting molecules with the rotational constant $B=0.3\,\hbar\omega$ in a one-dimensional harmonic trap in an external static electric field as a function of the isotropic interaction strength $g_0$ with the anisotropic interaction strength $g_{\pm 1}=4$ and electric field strength $d\mathcal{E}=0.5\,\hbar\omega$ and $2.5\,\hbar\omega$. These spectra result from the field-free spectrum shown in Fig.~\ref{fig:Evsgjj}(c). The assumed model of intermolecular interactions affects all states in the presence of a static electric field, because all eigenstates in this field are mixtures of field-free states with different total rotational angular momenta. For this reason, in Fig.~\ref{fig:Evsgjj+elf}, there is no state overlapping with the atomic result and the deviation from the atomic case is increasing with increasing the electric field strength. The electric field splits previously degenerate states into a larger number of states leading to a high density of states, especially in the strong electric field as plotted in Fig.~\ref{fig:Evsgjj+elf}(b).
The coupling between states originating from different total angular momenta results in the emergence of a large number of avoided crossings between these states when the electric field is applied. They are visible for all projections of the total rotational angular momentum along the field presented in Fig.~\ref{fig:Evsgjj+elf} and they are more pronounced for the larger electric field strength.   

Following adiabatically eigenstates across an avoided crossing can lead to a drastic (ex)change of eigenstates' properties.  
Figure~\ref{fig:J2vsg} shows mean values of the square of the total rotational angular momentum operator $\hat{\mathbf{J}}^2$ for selected eigenstates presented in Fig.~\ref{fig:Evsgjj+elf}(b). Selected eigenstates originate from states with $J=0-4$. It is apparent that each avoided crossing is associated with the exchange of the total rotational angular momenta between eigenstates. The widths of the avoided crossings depend on parameters of original states and the coupling strengths between them, but all observed transitions are smooth. Interestingly, such avoided crossing can be used to control and pump the total rotational angular momentum of the system by tuning intermolecular interactions or external electric field. Such a  control with an electric field would be a loose electric equivalent of using a magnetic field to pump the rotational angular momentum in the quantum variant of the Einstein--de Haas effect~\cite{Einstein15}.

\begin{figure}[tb!]
\begin{center}
\includegraphics[width=0.98\columnwidth]{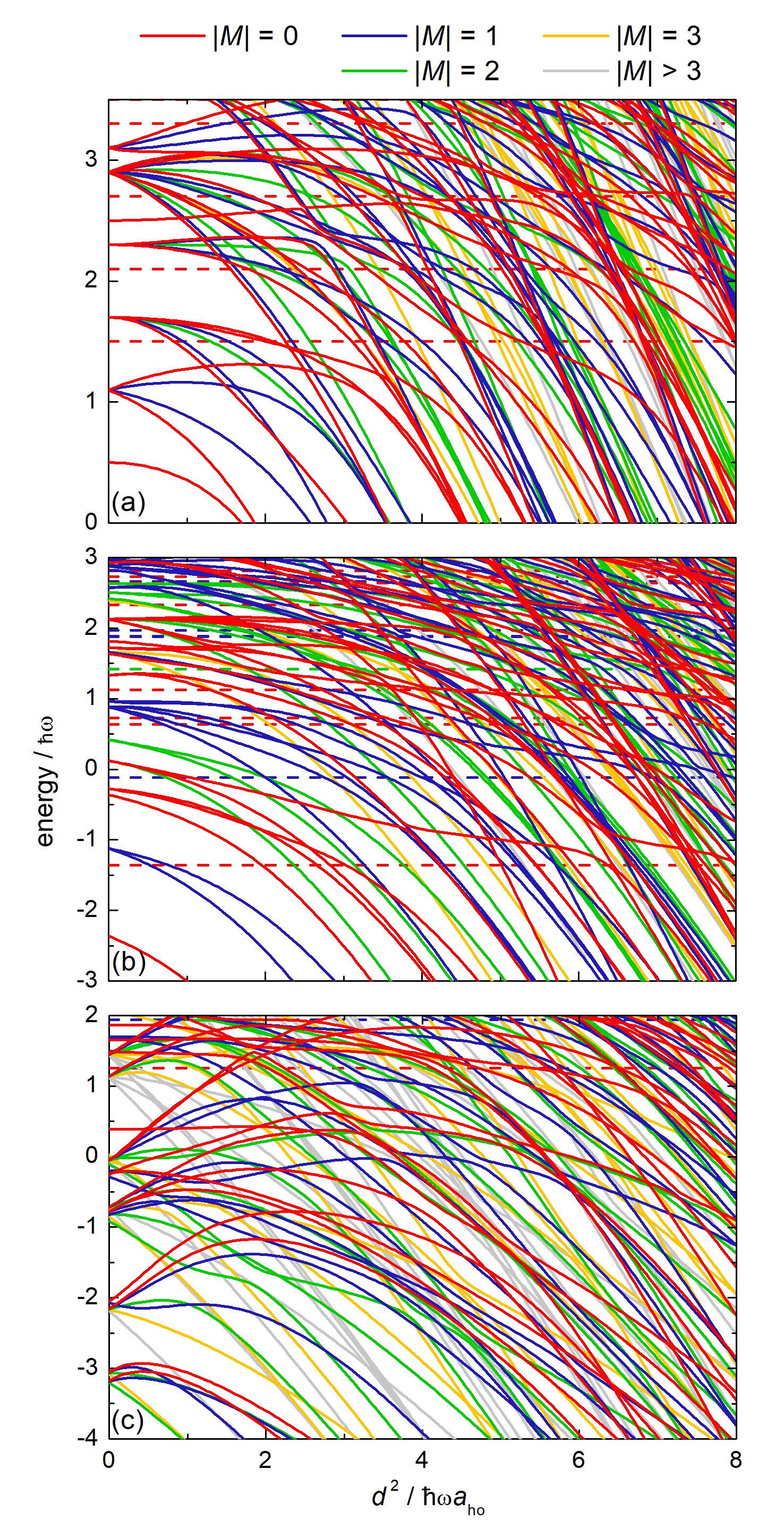}
\end{center}
\caption{Energy spectra of the relative motion for two interacting molecules with the rotational constant $B=0.3\,\hbar\omega$ in a one-dimensional harmonic trap in an external transverse or axial electric field as a function of the dipole-dipole interaction strength $d^2$: (a) $\mathcal{E}=0$, $g_0=0$, $g_{\pm 1}=0$, (b) $\mathcal{E}_z  = 2.5$, $g_0=0$, $g_{\pm 1}=0$, (c) $\mathcal{E}_z = 0.5$, $g_0=0$, $g_{\pm 1}=4$. Solid and dashed lines are for states of even and odd spatial symmetries. Different colors represent states with different projections $|M|$ of the total rotational angular momentum along the field.}
\label{fig:EvsD}
\end{figure}

Figure~\ref{fig:Evself} shows energy spectra of the relative motion for two interacting molecules with the rotational constant $B=0.3\,\hbar\omega$ in a one-dimensional harmonic trap in an external static electric field as a function of the electric field strength $d \mathcal{E}$ for different models of the isotropic and anisotropic short-range interactions. As a reference, panel~(a) presents the spectrum of the non-interacting molecules, thus it corresponds to the doubled result for a single polar molecule multiplied and shifted by trap vibrational energies only. Panels (b)-(d) show the spectra for the systems with the anisotropic interaction strength fixed at an intermediate value of $g_{\pm 1}=4$, whereas the isotropic interaction strength is zero in panel (b), intermediate in panel (c), and effectively infinite in panel (d). The electric field removes the degeneracy of states with different $|M|$ for a given $J$, leading to a high density of states especially when both the isotropic and anisotropic interaction strengths have intermediate values. Interestingly, for the system dominated by the anisotropic interaction presented in panel (b), the lowest states have high total rotational angular momentum, low energies, and relatively low density of states (in agreement with Fig.~\ref{fig:Evsgjk}(a)) and the effect of the electric field is weaker in such a scenario.   
In the limit of a very large isotropic interaction strength ($g_0\to\infty$), corresponding to the molecular Tonks-Girardeau limit, presented in panel (d), the importance of the anisotropic intermolecular interaction is reduced and the energy spectrum of the systems in the electric field simplifies to the spectrum of a single polar molecule in the field combed with the trap vibrational energies (spectra for odd and even spatial symmetries are the same). 

The complex spectra with numerous avoided crossings presented in Figs.~\ref{fig:Evsgjj+elf}-\ref{fig:Evself} raise a question whether investigated systems show a quantum chaotic behavior~\cite{Haake2010}. To verify this hypothesis we have calculated the nearest-neighbor spacing distributions of energy levels for investigated systems in a broad range of the intermolecular interactions and external fields strengths. In all calculations we have found level spacing distributions in much better agreement with the Poisson distribution than with the Wigner-Dyson one.
This observation strongly suggests that the statistical properties of calculated energy spectra do not follow the predictions of the Gaussian orthogonal ensemble of random matrices, and neither quantum chaotic behavior nor level repulsion is observed. The  investigated systems of two interacting molecules with short-range intermolecular interactions in the electric field behave thus rather like quantum integrable systems.   

\subsection{Dipole-dipole interaction}

The dipole-dipole interaction plays an important role in physics of ultracold molecules because heteronuclear molecules can possess a permanent electric dipole moment. At the same time, this interaction is of the long-range nature and can be controlled with external electric field~\cite{LemeshkoMP13}. In the present paper, we neglect its long-range character and focus on its impact on coupling and mixing molecular rotational angular momenta and interplay with an external static electric field.

\begin{figure}[tb!]
\begin{center}
\includegraphics[width=0.98\columnwidth]{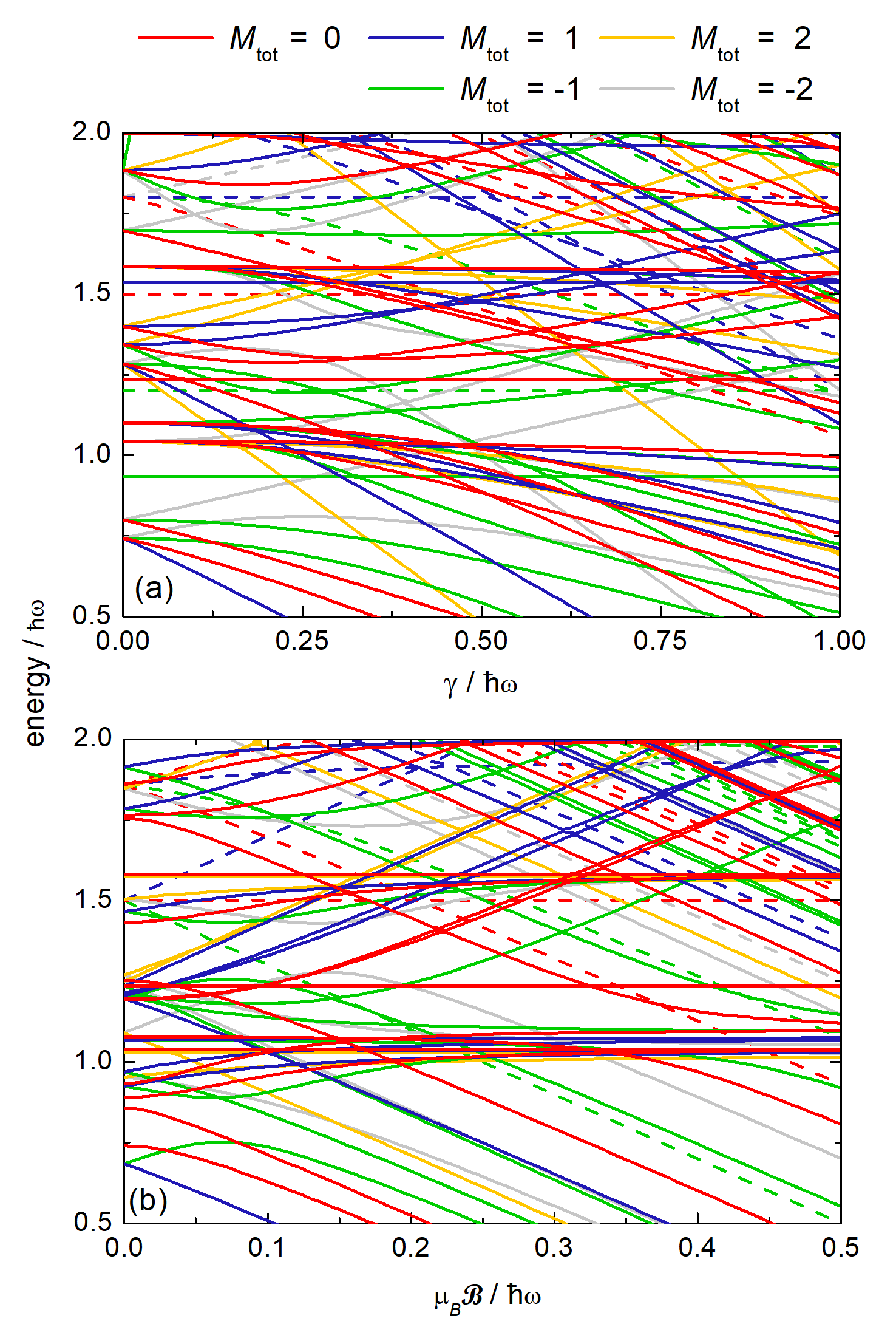}
\end{center}
\caption{Energy spectra of the relative motion for two interacting spin-1/2 molecules with the rotational constant $B=0.3\,\hbar\omega$ in a one-dimensional harmonic trap: (a) as a function of the spin-rotation coupling strength $\gamma$ with $g_0=4$, $g_{\pm 1}=4$, $\mathcal{B}=0.15\,\hbar\omega/\mu_B$ and (b) as a function of the magnetic field strength $\mathcal{B}$ with $g_0=4$, $g_{\pm 1}=4$, $\gamma=0.3\,\hbar\omega$. Basis set: $n_\text{max}=30$, $j_\text{max}=5$. Solid and dashed lines are for states of even and odd spatial symmetries. Different colors represent states with different projections $M$ of the total rotational angular momentum along the field.}
\label{fig:EvsA}
\end{figure}


Figure~\ref{fig:EvsD} presents the dependence of energy spectra of the relative motion for two interacting molecules with the rotational constant $B=0.3\,\hbar\omega$ in a one-dimensional harmonic trap in an external electric field as a function of the dipole-dipole interaction strength. Here, we define the dipole-dipole interaction strength as a square of the molecules' permanent electric dipole moment $d^2$. Panel~(a) shows the energy spectrum for the field-free case with the dipole-dipole interaction only. Panel~(b) presents the energy spectrum with the dipole-dipole interaction in a strong external electric field parallel to the trap axis. Panel~(c) presents the energy spectrum with the dipole-dipole interaction and additionally with the anisotropic interaction of intermediate strength also in an intermediate external electric field parallel to the trap axis. 

Numerous avoided crossings and high density of states are visible for all cases. The dipole-dipole interaction combines effects of the isotropic $g_0$ and anisotropic $g_{\pm 2}$ intermolecular interactions, thus it both shifts and splits energy levels. The dipole-dipole interaction restricted to one dimension does not conserve the total rotational angular momentum $J$, but its projection $M$ onto the trap axis is conserved, also in the presence of an electric field parallel to the trap axis. If an electric field is not parallel to the trap axis, then $M$ is not a good quantum number anymore. If the electric field is perpendicular to the trap axis, then it tends to align molecules perpendicularly to the trap axis leading to repulsive interaction between molecules and increasing their energy. No signature of quantum chaos is found in these spectra.

\subsection{Spin-rotation interaction and impact of external magnetic field}

If considered molecules posses a non-zero electronic spin angular momentum, then a static magnetic field can be used as a knob to control their interactions and dynamics in a trap via Zeeman effect~\cite{LemeshkoMP13}. In the present model, we assume that the intermolecular interaction potential does not depend on the electronic spin. Therefore, the magnetic field can couple with intermolecular dynamics through the molecular spin-rotation coupling only.

Panel~(a) in Fig.~\ref{fig:EvsA} presents the dependence of the energy spectrum of the relative motion for two interacting spin-1/2 molecules with the rotational constant $B=0.3\,\hbar\omega$ in a one-dimensional harmonic trap  on the spin-rotation coupling constant $\gamma$. Intermediate strengths of the isotropic and anisotropic interactions are assumed, $g_0=g_{\pm 1}=4$ together with the magnetic field of intermediate value $\mathcal{B}=0.15\,\hbar\omega/\mu_B$. Without magnetic field the total angular momentum $J_\text{tot},M_\text{tot}$ is conserved and real crossings between states with different $J_\text{tot},M_\text{tot}$ are expected. The energy spectrum gets more complex with increasing the spin-rotation coupling strength because the degeneracy of states related to the spin configuration is removed by this coupling. The magnetic field couples and splits states with different $J_\text{tot}$, while its projection $M_\text{tot}$ is conserved. Additionally, some crossings from the field-free case become avoided crossings in the magnetic field.

Panel~(b) in Fig.~\ref{fig:EvsA} shows the dependence of the energy spectrum of the relative motion for two interacting spin-1/2 molecules in a one-dimensional harmonic trap on the magnetic field strength $\mathcal{B}$. The intermolecular interactions are the same as in panel~(a) and an intermediate value of the spin-rotation coupling constant $\gamma=0.3\,\hbar\omega$ is assumed. If there is no spin-rotation coupling present in the system, then the magnetic field only simply splits and shifts states with different projections of the total electronic spin angular momentum on the magnetic field $M_S=m_{s_1}+m_{s_2}$, but the energy spectra for given $M_S$ look the same. However, when the rotational and spin angular momenta are coupled and mixed by the molecular spin-rotation coupling, then the magnetic field affects the system's dynamics and can be used to control it. The magnetic field induces numerous avoided crossings similarly as the electric field in previous sections. No signature of quantum chaos is found for spectra in the magnetic field, either.

\section{Summary and conclusions}
\label{sec:summary}

Motivated by experimental possibilities and ongoing efforts aiming at the production and application of fully controllable systems of few ultracold molecules trapped in optical tweezers or optical lattices, we have developed the model description of two interacting ultracold polar molecules effectively trapped in a one-dimensional harmonic potential. Molecules are described as distinguishable quantum rigid rotors interacting via multichannel two-body contact potential incorporating the short-range anisotropy of intermolecular interactions including dipole-dipole interaction. The form of the employed multichannel potential is motivated by the known nature of short-range chemical intermolecular interactions. We have also included interactions with external electric and magnetic fields via Stark and Zeeman effects, respectively. We have focused on systems with small rotational constants $B\leq\omega$, whereas detailed results for $B\gg \omega$ will be presented elsewhere.

We have carefully applied several approximations needed to simplify calculations, and to separate the impact and importance of different features of the molecular structure and intermolecular interactions on the system's dynamics. Thus, we have attempted to understand two interacting ultracold polar molecules trapped in a one-dimensional harmonic potential in a step-by-step manner. 
We have investigated the properties of such a system in a broad range of system parameters and external field strengths. We have analyzed in detail the interplay of the molecular rotational structure, anisotropic interactions, spin-rotation coupling, electric and magnetic fields, and harmonic trapping potential.

Our most important findings can be summarized as follows.
\begin{enumerate}
\item The anisotropic intermolecular interaction brings states with higher total rotational angular momenta to lower energies such that the absolute ground state of the molecular system can have the total angular momentum larger than zero and be degenerate. 
 
\item If the anisotropic interaction strength is larger than the isotropic one, then some energy levels for higher total rotational angular momenta diverge to minus infinity with increasing the intermolecular interaction strength. This indicates the emergence of the molecular equivalent of
the atomic super-Tonks-Girardeau limit with clustered ground state and excited gas-like super-Tonks states.


\item In the limit of a very large isotropic interaction strength, corresponding to the Tonks-Girardeau limit in the atomic system, the molecular character of the system is less pronounced and impact of the anisotropic interaction and electric field is smaller.

\item The electric and magnetic fields efficiently couple and mix states with different total angular momenta and result in complex energy spectra with a high density of states.

\item The electric and magnetic fields as well as dipole-dipole interaction induce a large number of avoided crossings.

\item Driving adiabatically the system across above avoided crossings can be used to control its properties. Especially, the total rotational angular momentum can be pumped to the system in a loose electric equivalent of the quantum Einstein--de Haas effect.

\item We have not found signatures of quantum chaotic behavior in energy spectra of investigated systems which suggests their quantum integrability.
\end{enumerate}

Replacing atoms with molecules in ultracold quantum few- and many-body systems opens up new possibilities stemming from molecules' rich internal structure and anisotropic intermolecular interactions, including long-range ones. 
Therefore, the present model and results may provide understanding and microscopic parameters for effective molecular many-body Hamiltonians. For example, our calculations may be considered as a microscopic model for the on-site interaction of the molecular multichannel Hubbard Hamiltonian. Thus, our results may be useful for the development of bottom-up molecule-by-molecule assembled molecular quantum simulators.

We believe that the results presented here will be followed by many applications of the proposed model and numerical approach to investigate interesting physics in different molecular systems, geometries, and dimensions. We foresee several possible extensions of the present paper. The most straightforward direction is moving to 2D or 3D arrangements. In contrast to the atomic case, no direct correspondence between energy spectra in one dimension and three dimensions is expected for the molecular system. The bosonic or fermionic nature of interacting molecules can also be addressed. On the other hand, the long-range character of the dipole-dipole interaction can be included and its interplay with isotropic and anisotropic short-range van der Waals interactions can be investigated.  Polar and paramagnetic molecules possessing at the same time both permanent electric dipole moment and spin structure can be studied together with their control with external electric and magnetic fields. Such systems may show an interesting interplay of magnetic and electric properties coupled by the molecular internal structure. Hyperfine structure of molecules and chemical reactivity can also be included. Molecules possessing different masses, rotational constants, and trapping frequencies, and resulting coupling between the center-of-mass and relative motions, can be considered. Emergence of quantum chaotic properties with increasing complexity of the system is another intriguing question. Time-dependent dynamics in few-body molecular systems, especially after a quench of system parameters, is another not explored but potentially interesting direction of research. Finally, two interacting molecules can be trapped in two sites of an optical lattice or optical tweezer and such a double-well configuration can be investigated as a fundamental building block for the implementation of quantum gates and quantum computation with molecular systems. 

\begin{acknowledgments}
We would like to thank Grzegorz Cha{\l}asinski, Zbigniew Idziaszek, and Piotr \.Zuchowski for useful discussions. We acknowledge financial support from the Foundation for Polish Science within the Homing program co-financed by the European Union under the European Regional Development Fund and the PL-Grid Infrastructure. M.~L. acknowledges the Spanish Ministry MINECO (National Plan 15 FISICATEAMO Grant No. FIS2016-79508-P and SEVERO OCHOA Grant No. SEV-2015-0522); Fundaci\'o Cellex, Generalitat de Catalunya (AGAUR Grant No. 2017 SGR 1341 and CERCA/Program); European Research Council AdG OSYRIS; EU FETPRO QUIC; and the National Science Centre, Poland-Symfonia Grant No. 2016/20/W/ST4/00314.  
\end{acknowledgments}

\onecolumngrid

\appendix
\section{Matrix elements}
\label{sec:app}

Here, we provide matrix elements of the components of the Hamiltonian given by Eq.~\eqref{eq:relHam} defined in Eqs.~\eqref{eq:Ham_rot}, \eqref{eq:Ham_field}, \eqref{eq:Hamiso}, \eqref{eq:Hamaniso}, \eqref{eq:Hamdip} in the computation basis of $|n,J,M,j_1,j_2,s_1,m_{s_1},s_2,m_{s_2}\rangle\equiv|n\rangle|J,M,j_1,j_2\rangle |s_1,m_{s_1}\rangle|s_2,m_{s_2}\rangle$ as described in Sec.~\ref{sec:theory},
\begin{equation} 
\langle \hat{H}_X \rangle \equiv \langle n,J,M,j_1,j_2,s_1,m_{s_1},s_2,m_{s_2} | \hat{H}_X |n',J',M',j_1',j_2',s_1',m_{s_1}',s_2',m_{s_2}' \rangle.
\end{equation}
\begin{equation} 
\langle \hat{H}_{\text{trap}} \rangle = \delta_{n n'} \delta_{J J'} \delta_{M M'} \delta_{j_1 j_1'} \delta_{j_2 j_2'} \delta_{s_1 s_1'} \delta_{s_2 s_2'} \delta_{m_{s_1} m_{s_1}'} \delta_{m_{s_2} m_{s_2}'}\hbar\omega \left(n + \frac{1}{2}\right),
\end{equation} 
\begin{equation} 
\langle \hat{H}_{\text{rot}} \rangle = \delta_{n n'} \delta_{J J'} \delta_{M M'} \delta_{j_1 j_1'} \delta_{j_2 j_2'} \delta_{s_1 s_1'} \delta_{s_2 s_2'} \delta_{m_{s_1} m_{s_1}'} \delta_{m_{s_2} m_{s_2}'} B\left(j_1(j_1 + 1) + j_2(j_2 + 1)\right),
\end{equation} 
\begin{equation}
\begin{gathered}
\langle \hat{H}_{\text{spin--rot}} \rangle = \delta_{n n'} \delta_{J J'} \delta_{M + m_{s_1} + m_{s_2}, M' + m_{s_1}' + m_{s_2}'} \delta_{j_1 j_1'} \delta_{j_2 j_2'} \delta_{s_1 s_1'} \delta_{s_2 s_2'} \\
\times\sum_{m_1 = -j_1}^{j_1} \sum_{m_2 = -j_2}^{j_2} \sum_{m_1' = -j_1}^{j_1} \sum_{m_2' = -j_2}^{j_2} \braket{j_1 m_1' j_2 m_2'}{J M} \braket{j_1 m_1' j_2 m_2'}{J M'} \\
\times\left(\delta_{m_{s_1} m_{s_1}'} \delta_{m_{s_2} m_{s_2}'} \gamma (m_1 m_{s_1} + m_2 m_{s_2}) \right. \\
+ \frac{\gamma}{2}\left(\delta_{m_{s_1}+1,m_{s_1}'} \delta_{m_{s_2},m_{s_2}'} \delta_{m_1-1,m_1'} \delta_{m_2 m_2'} + \delta_{m_{s_1},m_{s_1}'} \delta_{m_{s_2}+1,m_{s_2}'} \delta_{m_1 m_1'} \delta_{m_2 - 1,m_2'} \right.\\
\left.\left.+ \delta_{m_{s_1}-1,m_{s_1}'} \delta_{m_{s_2},m_{s_2}'} \delta_{m_1+1,m_1'} \delta_{m_2,m_2'} + \delta_{m_{s_1},m_{s_1}'} \delta_{m_{s_2}-1,m_{s_2}'} \delta_{m_1 m_1'} \delta_{m_2 + 1,m_2'} \right)\right)\,,
\end{gathered}
\end{equation}
\begin{equation}
\begin{gathered}
\langle \hat{H}_{\text{Stark}}(\mathcal{E}_z\neq 0) \rangle = - d\mathcal{E}_z \delta_{n n'} \delta_{s_1 s_1'} \delta_{s_2 s_2'} \delta_{m_{s_1} m_{s_1}'} \delta_{m_{s_2} m_{s_2}'} \\
\times\sum_{m_1 = -j_1}^{j_1} \sum_{m_2 = -j_2}^{j_2} \sum_{m_1' = -j_1'}^{j_1'} \sum_{m_2' = -j_2'}^{j_2'} \braket{j_1 m_1 j_2 m_2}{J M} \braket{j_1' m_1' j_2' m_2'}{J' M'}\\
\times\delta_{m_1 m_1'} \delta_{m_2 m_2'} \left(\delta_{j_1 \pm 1, j_1'} \delta_{j_2 j_2'} \sqrt{\frac{2j_1+1}{2j_1'+1}} \bra{j_1 0 1 0}\ket{j_1' 0} \bra{j_1 m_1 1 0}\ket{j_1' m_1} + \right.\\
\left. + \delta_{j_1 j_1'} \delta_{j_2 \pm 1, j_2'} \sqrt{\frac{2j_2+1}{2j_2'+1}} \bra{j_2 0 1 0}\ket{j_2' 0} \bra{j_2 m_2 1 0}\ket{j_2' m_2}\right)
\end{gathered}
\end{equation}
\begin{equation}
\begin{gathered}
\langle \hat{H}_{\text{Stark}}(\mathcal{E}_x\neq 0) \rangle = \frac{d\mathcal{E}_x}{\sqrt{2}} \delta_{n n'} \delta_{s_1 s_1'} \delta_{s_2 s_2'} \delta_{m_{s_1} m_{s_1}'} \delta_{m_{s_2} m_{s_2}'} \\
\times\sum_{m_1 = -j_1}^{j_1} \sum_{m_2 = -j_2}^{j_2} \sum_{m_1' = -j_1'}^{j_1'} \sum_{m_2' = -j_2'}^{j_2'} \braket{j_1 m_1 j_2 m_2}{J M} \braket{j_1' m_1' j_2' m_2'}{J' M'}\\
\times\left(\delta_{j_1 \pm 1, j_1'} \delta_{j_2 j_2'} \delta_{m_2 m_2'} \sqrt{\frac{2j_1+1}{2j_1'+1}} \bra{j_1 0 1 0}\ket{j_1' 0}
\left(\delta_{m_1 + 1, m_1'} \bra{j_1 m_1 1 1}\ket{j_1' m_1'} - \delta_{m_1 - 1, m_1'} \bra{j_1 m_1 1 (-1)}\ket{j_1' m_1'}\right) \right. \\
\left.+ \delta_{j_1 j_1'} \delta_{j_2 \pm 1, j_2'} \delta_{m_1 m_1'} \sqrt{\frac{2j_2+1}{2j_2'+1}} \bra{j_2 0 1 0}\ket{j_2' 0}
\left(\delta_{m_2 + 1, m_2'} \bra{j_2 m_2 1 1}\ket{j_2' m_2'} - \delta_{m_2-1, m_2'} \bra{j_2 m_2 1 (-1)}\ket{j_2' m_2'}\right)\right)\,,
\end{gathered}
\end{equation}
\begin{equation}
\langle \hat{H}_{\text{Zeeman}} \rangle = \delta_{n n'} \delta_{J J'} \delta_{M M'} \delta_{j_1 j_1'} \delta_{j_2 j_2'} \delta_{s_1 s_1'} \delta_{s_2 s_2'} \delta_{m_{s_1} m_{s_1}'} \delta_{m_{s_2} m_{s_2}'} 2 \mu_B (m_{s_1} + m_{s_2}) \mathcal{B}\,,
\end{equation}
\begin{equation} 
\langle \hat{H}_{\text{iso}} \rangle = \delta_{J J'} \delta_{M M'} \delta_{j_1 j_1'} \delta_{j_2 j_2'} \delta_{s_1 s_1'} \delta_{s_2 s_2'} \delta_{m_{s_1} m_{s_1}'} \delta_{m_{s_2} m_{s_2}'} g_0 \varphi_{n} (0) \varphi_{n'}(0)\,,
\end{equation} 
\begin{equation}
\langle \hat{H}_{\text{aniso}} \rangle = \delta_{J J'} \delta_{M M'} \delta_{j_1 \pm k, j_1'} \delta_{j_2 \mp k, j_2'} \delta_{s_1 s_1'} \delta_{s_2 s_2'} \delta_{m_{s_1} m_{s_1}'} \delta_{m_{s_2} m_{s_2}'} g_{\pm k}\varphi_{n} (0) \varphi_{n'}(0)\,,
\end{equation}
\begin{equation}
\begin{gathered}
\langle \hat{H}_{\text{dip}} \rangle = - d^2 \varphi_{n} (0) \varphi_{n'} (0) \delta_{s_1 s_1'} \delta_{s_2 s_2'} \delta_{m_{s_1} m_{s_1}'} \delta_{m_{s_2} m_{s_2}'} \delta_{j_1 \pm 1, j_1'} \delta_{j_2 \pm 1, j_2'} \\
\times\sqrt{\left(\frac{2j_1+1}{2j_1'+1}\right)\left(\frac{2j_2+1}{2j_2'+1}\right)} \braket{j_1 0 1 0}{j_1' 0} \braket{j_2 0 1 0}{j_2' 0}\\
\times\sum_{m_1 = -j_1}^{j_1} \sum_{m_2 = -j_2}^{j_2} \sum_{m_1' = -j_1'}^{j_1'} \sum_{m_2' = -j_2'}^{j_2'} \left(\delta_{m_1 + 1,m_1'} \delta_{m_2 - 1,m_2'} \braket{j_1 m_1 1 1}{j_1' m_1'} \braket{j_2 m_2 1 (-1)}{j_2' m_2'}\right. \\
\left.+\delta_{m_1 - 1,m_1'} \delta_{m_2 + 1,m_2'} \braket{j_1 m_1 1 (-1)}{j_1' m_1'} \braket{j_2 m_2 1 1}{j_2' m_2'} + 2 \delta_{m_1 m_1'} \delta_{m_2 m_2'} \braket{j_1 m_1 1 0}{j_1' m_1} \braket{j_2 m_2 1 0}{j_2' m_2}\right)\,,
\end{gathered}
\end{equation}
where $\delta_{ij}$ is the Kronecker delta and $\varphi_{n}(0)$ is a harmonic oscillator wave function for state $n$ at point $z=0$.

\twocolumngrid

\bibliography{two_molecules_1D}

\end{document}